%% file: arxiv.tex
\pdfoutput=1
\documentclass{article}
\usepackage{iclr2027_conference,times}
\input{math_commands.tex}

\usepackage[utf8]{inputenc}
\usepackage[T1]{fontenc}
\usepackage{amsmath,amssymb,bm}
\usepackage{graphicx}
\usepackage{booktabs}
\usepackage{multirow}
\usepackage{subcaption}
\usepackage{xcolor}
\usepackage{enumitem}
\usepackage{microtype}
\usepackage{url}
\usepackage[hidelinks]{hyperref}
\usepackage{titletoc}

\graphicspath{{figs/}}
\setlist{itemsep=1pt,topsep=2pt,leftmargin=*}

\newif\ifdraftmode
\draftmodefalse
\ifdraftmode
  
  \newcommand{\note}[1]{\textcolor{red!70!black}{\small[\textit{#1}]}}
\else
  
  \newcommand{\note}[1]{}
\fi

\newcommand{\CD}{\ensuremath{C_D}}

\newcommand{\dCD}{\ensuremath{\Delta C_D}}
\newcommand{\relLtwo}{\ensuremath{\varepsilon_{L_2}}}

\newcommand{\pref}{\ensuremath{p_{\mathrm{ref}}}}
\newcommand{\Uinf}{\ensuremath{U_\infty}}
\newcommand{\Aref}{\ensuremath{A_{\mathrm{ref}}}}
\newcommand{\WSS}{\ensuremath{\tau_w}}
\newcommand{\yplus}{\ensuremath{y^{+}}}
\newcommand{\Rey}{\ensuremath{\mathrm{Re}}}

\newcommand{\SHIFTTruck}{SHIFT-Truck}
\newcommand{\SMART}{SMART}
\newcommand{\DoMINO}{DoMINO}
\newcommand{\GeoT}{GeoTransolver}
\newcommand{\ABUPT}{AB-UPT}

\input{numbers}

\def\ArxivVersion{}
\renewcommand{\nRelease}{1{,}000}
\renewcommand{\nGPUhCorpus}{24{,}000}   % 1,000 cases x 6 h x 4 H100
\newcommand{\ShiftTruckURL}{\url{https://huggingface.co/datasets/luminary-shift/Truck}}

\title{\SHIFTTruck{}: A High-Fidelity Aerodynamics Dataset and Benchmark for Pickup Trucks}

\author{Riddhiman Raut$^{1}$, Yin Yu$^{1}$, Aashwin Anand Mishra$^{1}$, Michael Emory$^{1}$, \\
\bfseries Thomas Economon$^{1}$, Peter Lyu$^{1}$ \& Juan J. Alonso$^{1,2}$ \\
$^{1}$Luminary, San Mateo, CA, USA \\
$^{2}$Department of Aeronautics and Astronautics, Stanford University, Stanford, CA, USA
}

\iclrfinalcopy

\begin{document}
\maketitle
\lhead{Preprint. Under review.}

\input{sections/00-abstract}
\input{sections/01-intro}
\input{sections/03-dataset}
\input{sections/04-benchmark}

\input{sections/05-results}
\input{sections/06-ood}
\input{sections/07-conclusion}

\bibliography{refs}
\bibliographystyle{iclr2027_conference}

\clearpage
\appendix
\section*{Appendix contents}
\startcontents[appendix]
{\setlength{\parskip}{0pt}\printcontents[appendix]{}{1}{\setcounter{tocdepth}{2}}}
\clearpage
\input{appendix/A-geometry}
\input{appendix/B-cfd}
\input{appendix/C-verification-validation}
\input{appendix/D-design-space}
\input{appendix/E-benchmark-details}
\input{appendix/F-release}
\input{appendix/G-datasets}

\end{document}

%% file: math_commands.tex
\usepackage{amsmath,amsfonts,bm}

\def\eqref#1{equation~\ref{#1}}
\def\1{\bm{1}}

\DeclareMathAlphabet{\mathsfit}{\encodingdefault}{\sfdefault}{m}{sl}
\SetMathAlphabet{\mathsfit}{bold}{\encodingdefault}{\sfdefault}{bx}{n}

%% file: numbers.tex
\newcommand{\nRelease}{800}          % [stated by Rik 2026-09-23] cases at publication (787 simulated as of 2026-09-23). The benchmark trains on \nCorpus.
\newcommand{\nCorpus}{712}           % [ledger] cases in the frozen benchmark corpus
\newcommand{\nSobolPoints}{1{,}000}  % [ledger] design points drawn
\newcommand{\nParams}{17}            % [ledger]
\newcommand{\nParamsCont}{15}        % [ledger]
\newcommand{\nParamsDisc}{2}         % [ledger]
\newcommand{\nCageVerts}{126}        % [v3]
\newcommand{\nBaseTrucks}{248}       % [v3]
\newcommand{\nBaseTrucksPlanned}{250} % [HF card] Sobol points, each crossed with the four cover/spoiler combinations
\newcommand{\nPhysTime}{6.25}       % [HF card] s of physical time per case (metadata.json)
\newcommand{\nAvgTimeCorpus}{2.0}   % [HF card] s, averaging window
\newcommand{\nBedPairs}{219}         % [ledger] trucks carrying both bed states
\newcommand{\nBedPairsMatched}{218}   % [csv 712] trucks with both bed states, chin spoiler held fixed
\newcommand{\nSpoilerPairs}{245}     % [v3]
\newcommand{\nClosed}{492}           % [v3]
\newcommand{\nOpen}{220}             % [v3]
\newcommand{\nBodyLenLo}{4.90}       % [v3] m
\newcommand{\nBodyLenHi}{5.87}       % [v3] m
\newcommand{\nHeightLo}{1.49}        % [v3] m
\newcommand{\nHeightHi}{1.90}        % [v3] m
\newcommand{\nWettedLo}{41.5}        % [v3] m^2
\newcommand{\nWettedHi}{53.7}        % [v3] m^2

\newcommand{\nTest}{90}              % [ledger]
\newcommand{\nTestTrucks}{30}        % [ledger] base trucks in test
\newcommand{\nVal}{72}             % [split] confirmed 2026-09-22
\newcommand{\nTrain}{550}          % [split] confirmed 2026-09-22
\newcommand{\nSplitNNTest}{1.87}     % [csv 712 + Sobol seed 42 regen, 2026-09-25] median NN distance test -> train, normalized [-1,1]^15
\newcommand{\nSplitNNTrain}{1.89}    % median leave-one-out NN distance within train
\newcommand{\nSplitNNp}{0.86}        % KS p, test NN vs train LOO NN
\newcommand{\nSplitMargp}{0.06}      % smallest KS p over the 15 marginals, train vs test
\newcommand{\nSplitCDTestLo}{0.277}
\newcommand{\nSplitCDTestHi}{0.549}
\newcommand{\nSplitCDTestMed}{0.400}
\newcommand{\nSplitCDTrainMed}{0.383}

\newcommand{\nCellsBase}{122}        % [ledger] million cells, baseline GTU
\newcommand{\nCellsCorpus}{100}      % [v3] million cells, typical morphed case (97.6 M for variant 0000)
\newcommand{\nSteps}{20{,}000}       % [ledger]
\newcommand{\nAvgStepsCorpus}{10{,}000} % [v3] averaging window, corpus runs
\newcommand{\nAvgCTUCorpus}{15}      % [HF card] 10,000 steps x 2.0e-4 s = 2.0 s = 14.8 CTU
\newcommand{\nAvgStepsVal}{18{,}000} % [ledger] averaging window, validation runs
\newcommand{\nDt}{2.0\times10^{-4}}  % [HF card] s
\newcommand{\nRANSIters}{4{,}000}    % [ledger]
\newcommand{\nBLLayers}{5}           % [ledger]
\newcommand{\nBLER}{1.2}             % [ledger]
\newcommand{\nYplusLo}{20}           % [ledger]
\newcommand{\nYplusHi}{35}           % [ledger]
\newcommand{\nYplusFineLo}{2}        % [v3] mirrors and chin spoiler
\newcommand{\nYplusFineHi}{4}        % [v3]
\newcommand{\nBelts}{5}              % [ledger]
\newcommand{\nRefLen}{5.25}          % [ledger] m
\newcommand{\nRefArea}{2.65}         % [ledger] m^2
\newcommand{\nUinf}{38.89}           % [ledger] m/s
\newcommand{\nUinfKmh}{140}          % [v3] km/h
\newcommand{\nPref}{101{,}325}       % [ledger] Pa
\newcommand{\nTref}{288.15}          % [HF card] K, reference temperature of the force/pressure coefficients (params.json ref_t)
\newcommand{\nTstatic}{293.15}       % [HF card; Rik 2026-09-24] K, freestream static temperature
\newcommand{\nRhoFree}{1.204}        % [derived] kg/m^3, 101325 / (287.058 x 293.15)
\newcommand{\nRho}{1.225}            % [derived] kg/m^3, reference density of the coefficients, 101325 / (287.058 x 288.15)
\newcommand{\nRe}{1.4\times10^{7}}   % [HF card ~1.4e7] 1.36e7 at 293.15 K with Sutherland viscosity; quoted to two figures
\newcommand{\nCTU}{0.135}            % [v3] s
\newcommand{\nDomX}{59}              % [v3] m
\newcommand{\nDomY}{50}
\newcommand{\nDomZ}{30}
\newcommand{\nBlockage}{0.18}        % [v3] per cent
\newcommand{\nGPUsPerCase}{4}        % [ledger]
\newcommand{\nGPUsVal}{8}            % [ledger]
\newcommand{\nWallClockVal}{7\,h\,02\,min} % [ledger] on 8 H100
\newcommand{\nGPUsScale}{56}         % [ledger]
\newcommand{\nWallClockScale}{1\,h\,27\,min} % [ledger] on 56 H100
\newcommand{\nWallClockCorpus}{6\,h}      % [stated by Rik 2026-09-23] wall clock per corpus case on \nGPUsPerCase H100
\newcommand{\nGPUhCorpus}{19{,}200}   % [derived] \nRelease x 6 h x 4 H100; update if \nRelease changes

\newcommand{\nGridLevels}{5}         % [v3]
\newcommand{\nGridCells}{26, 80, 107, 113 and 304} % [v3] million
\newcommand{\nGridCDa}{0.443}        % [v3]

\newcommand{\nGridCDe}{0.427}
\newcommand{\nBlendSens}{0.003}      % [v3] mirror-delta sensitivity to blending
\newcommand{\nRepeatCDLo}{0.07}      % [v3] per cent
\newcommand{\nRepeatCDHi}{0.7}       % [v3] per cent
\newcommand{\nBaseCD}{0.446}         % [ledger]
\newcommand{\nExpBaseCD}{0.393}      % [ledger]
\newcommand{\nAbsOffsetCD}{0.053}    % [ledger]
\newcommand{\nDeltaMirrorLum}{-0.0106}
\newcommand{\nDeltaMirrorExp}{-0.012}
\newcommand{\nDeltaMirrorRef}{-0.010}
\newcommand{\nDeltaAirdamLum}{+0.0349}
\newcommand{\nDeltaAirdamExp}{+0.028}
\newcommand{\nDeltaAirdamRef}{+0.035}
\newcommand{\nValTol}{0.007}         % [ledger] larger of the two delta discrepancies

\newcommand{\nCDSd}{0.047}           % [v3]
\newcommand{\nCDMedClosed}{0.372}    % [v3]
\newcommand{\nCDMedOpen}{0.412}      % [v3]
\newcommand{\nRhoHeightClosed}{0.53} % [v3] Spearman
\newcommand{\nRhoHeightOpen}{0.60}
\newcommand{\nBedMedian}{+0.018}     % [csv 712] like for like: open vs closed, chin spoiler present in both, 218 trucks
\newcommand{\nBedMin}{+0.005}
\newcommand{\nBedMax}{+0.050}
\newcommand{\nSpoilerMedian}{0.036}  % [v3]
\newcommand{\nLandAhmedML}{500}
\newcommand{\nLandWindsorML}{355}
\newcommand{\nLandDrivAerML}{500}
\newcommand{\nLandDrivAerNet}{4{,}000}
\newcommand{\nLandDrivAerNetPP}{8{,}000}
\newcommand{\nLandDrivAerStar}{12{,}000}
\newcommand{\nLandSHIFTSUV}{2{,}000}   % [stated by Rik 2026-09-25] cases in the release as of September 2026; release still growing
\newcommand{\nLightTruckShare}{66}   % [epa2025trends] per cent of MY2024 production classed as trucks under NHTSA regulations
\newcommand{\nPickupShare}{14}       % [epa2025trends] per cent of MY2024 production (14.1); was 15 in MY2023 (EPA-420-R-24-022)

\newcommand{\nTrainEpochs}{500}    % [author correction, 2026-09-26] all models and tracks; FP32
\newcommand{\nSeedSdGeoT}{0.002}       % [stated by Rik 2026-09-25] s.d. of drag R^2 over seeds 42, 43, 44
\newcommand{\nSeedSdABUPT}{0.010}
\newcommand{\nSeedSdSMART}{0.010}
\newcommand{\nSeedSdDoMINO}{0.034}      % 0.0339
\newcommand{\nInfABUPT}{883}     % [stated by Rik 2026-09-25] s per case, surface track, 8.8 M faces, one T4, fp32
\newcommand{\nInfGeoT}{630}
\newcommand{\nInfSMART}{267}
\newcommand{\nInfDoMINO}{102}
\newcommand{\nInfFaces}{8.8}      % million faces of the inference surface
\newcommand{\nSgeoCpPct}{2}         % GeoTransolver surface Cp MAE 0.021 = 2% of q_inf (926.4 Pa)
\newcommand{\nCDRsqLeadLo}{0.95}    % lowest drag R^2 of GeoTransolver/AB-UPT is 0.953
\newcommand{\nSsmartCDMAE}{0.005}   % SMART surface drag MAE 0.00465
\newcommand{\nSsmartDragN}{11.4}    % 0.00465 * q_inf (926.4 Pa) * A_ref (2.65 m^2) = 11.4 N

%% file: sections/00-abstract.tex
% Abstract -- as registered on OpenReview (2026-09-18). Edit with care: the
% registered abstract may be revised at the full-paper deadline, but the claims
% here are the ones the rest of the paper has to carry.
\begin{abstract}
Pickup trucks account for \nPickupShare{}\% of new light-duty vehicles produced in the United States and yet, are among the least aerodynamic. Their open cargo bed adds a flow that existing automotive aerodynamics datasets such as DrivAerML and SHIFT-SUV do not contain, in which the shear layer leaving the cab roof passes over a recirculating bed flow before separating again at the tailgate. The resulting high aerodynamic drag on pickup trucks decreases fuel efficiency while increasing emissions in internal combustion engines, while also limiting highway range on electric counterparts. Drag reduction through design optimization is therefore crucial, however, the time and compute required to run scale-resolved Computational Fluid Dynamics (CFD) simulations serve as a bottleneck to expansive design exploration. Neural surrogates offer a way to predict flow features at a fraction of that cost, provided they are trained on large-scale, high-fidelity and domain specific data. To bridge this gap, we introduce \SHIFTTruck{}, the first such dataset for pickup trucks. It comprises \nRelease{} Spalart--Allmaras delayed detached-eddy simulations (SA-DDES) of a reference pickup geometry morphed across 17 shape parameters, such as cab geometry, bed dimensions and windshield rake. Every case is computed on a mesh of about \nCellsCorpus{} million cells at a Reynolds number of $\nRe$ and released with time-averaged surface pressure, wall shear stress, volumetric pressure and velocity. The numerical setup is verified by grid refinement and repeated simulations, as well as checked against wind-tunnel measurements. We define geometry-grouped training, validation and test splits and benchmark four recent neural surrogate architectures, \DoMINO{}, \GeoT{}, \ABUPT{} and \SMART{}, on surface and volume tracks. Beyond in-distribution evaluation, \SHIFTTruck{} introduces controlled distribution shifts in the operating point, the discretization of the input surface and the vehicle archetype. We find that models with strong in-distribution performance can exhibit substantial degradation under distribution shifts: operating-condition changes expose failures to infer speed dependence, while tessellation and cross-vehicle shifts reveal markedly different robustness across architectures. These controlled evaluations make SHIFT-Truck a benchmark not only for aerodynamic surrogate accuracy, but also for studying generalization of neural surrogates across physical and numerical distributions.
\end{abstract}

%% file: sections/01-intro.tex
\section{Introduction}
\label{sec:intro}

Light-duty trucks make up \nLightTruckShare{}\% of new light-duty vehicles produced in the United States, and pickup trucks alone account for \nPickupShare{}\%~\citep{epa2025trends}. Despite their prevalence, pickup trucks are aerodynamically poor, owing to a bluff front with a large cross-sectional area, a high ground clearance that admits turbulent flow underneath, and an open cargo bed that causes flow separation and a large turbulent wake. At highway speeds aerodynamic drag is the dominant resistive force. Modern sedans have drag coefficients of $0.2$ to $0.25$ \citep{hucho2013aerodynamics, heft2012drivaer}, while pickup trucks range from $0.45$ to $0.5$ \citep{cfd_pickup_drag}, representing an increase of roughly $60\%$ to $100\%$ in aerodynamic resistance. For Internal Combustion Engine (ICE) fleets, this means higher fuel consumption and emissions, and a reduction of $10$ to $15\%$ in drag at highway speed is estimated to save $5$ to $7\%$ of fuel~\citep{bellman2010reducing}. For electric vehicles (EVs), the same penalty dictates heavier batteries and limits highway range. For ICE and EV pickups alike, aerodynamic optimization of their design is a critical engineering need.

To optimize vehicle aerodynamics, engineers need to explore a large, high-dimensional design space with Computational Fluid Dynamics (CFD) over hundreds of design iterations. However, resolving the turbulent flow with scale-resolving approaches such as Detached Eddy Simulation (DES) on massive meshes takes hours to even days per design on high-performance computing clusters, making design exploration infeasible. Steady Reynolds-Averaged Navier-Stokes (RANS) simulations cost less but are less reliable in such flows dominated by separation, recirculation and streamline curvature~\citep{ashton2015key, ashton2016assessment}. In recent years, neural surrogate models such as \DoMINO{}~\citep{domino2025} and \GeoT{}~\citep{geotransolver2025} have emerged as an alternative to traditional CFD. By learning a mapping from raw geometry and operating conditions to output states, they bypass the iterative solver and predict complex flow fields in minutes on a single GPU. Their accuracy, however, is bounded by the scale, fidelity and domain of their training data, restricting widespread practical usage of these models.

While the Scientific Machine Learning community has benefited from several large-scale automotive CFD datasets such as AhmedML~\citep{ashton2024ahmedml}, WindsorML~\citep{ashton2024windsorml}, DrivAerML~\citep{ashton2024drivaerml}, DrivAerNet~\citep{elrefaie2024drivaernet}, DrivAerNet++~\citep{elrefaie2024drivaernetpp}, DrivAerStar~\citep{qiu2025drivaerstar} and SHIFT-SUV~\citep{shift_placeholder} (Table~\ref{tab:landscape}), these datasets have typically focused on closed passenger vehicles. Pickups introduce a distinct aerodynamic challenge through their stepped cab and bed geometry. With a flush bed cover, flow separates behind the cab above the bed, while another wake forms behind the tailgate. Opening the bed exposes a cavity that further changes the recirculating flow and its interaction with the rear wake. Cab height, bed length, and bed configuration therefore influence the flow around the entire rear of the vehicle, making their effects on drag important targets for aerodynamic optimization. Capturing these relationships requires training and evaluation data that span a wide range of pickup geometries and include both open and covered beds.

To address this need, we introduce \SHIFTTruck{}, to our knowledge the first large-scale, high-fidelity external aerodynamics dataset for pickup trucks. It comprises \nRelease{} Spalart--Allmaras Delayed Detached-Eddy Simulations (SA-DDES) of variants of the open Generic Truck Utility (GTU) geometry~\citep{woodiga2020gtu}, morphed across 15 continous design parameters that capture cab geometry, windshield rake, bed dimensions and ride height. Also, the bed cover and chin spoiler serve as two features that change the wetted surface, and can be turned on and off. Every dataset case is computed on a mesh of about \nCellsCorpus{} million cells at a Reynolds number of $\nRe$ and carries time-averaged surface pressure, wall shear stress, volumetric pressure and velocity. The numerical setup is checked against wind-tunnel measurements on the same truck.

To establish \SHIFTTruck{} as a benchmark, we define geometry-grouped train, validation and test splits and evaluate four recent neural surrogate architectures, \DoMINO{}, \GeoT{}, \ABUPT{} and \SMART{}, on separate surface and volume tracks. Strong in-distribution accuracy, however, does not show whether a surrogate has learned the physical mapping from geometry to flow or only features tied to its training distribution. \SHIFTTruck{} makes this distinction measurable through Out-Of-Distribution sets with controlled shifts in the operating point with the geometry held fixed, in the discretization of the input surface with the geometry and the flow held fixed and in the vehicle archetype, from pickup to SUV. On held-out In-Distribution trucks, three of the four models recover drag with $R^2$ between 0.95 and 0.97. However, under the distribution shifts, their behavior diverges: none of the models captures a change in freestream speed, their sensitivity to surface decimation differs by more than an order of magnitude and all four models underpredict drag on SUVs. In summary, the primary contributions of this work are as follows:
\begin{itemize}
  \item \textbf{Dataset.} A \nParams{}-parameter design space (\nParamsCont{} continuous morphs of cab, bed, nose and ride height, and \nParamsDisc{} discrete parts, a bed cover and a chin spoiler), sampled by a Sobol sequence and solved with SA-DDES, with time-averaged surface and volume fields for every case (\S\ref{sec:dataset}).
    \item \textbf{CFD assessment.} Grid-refinement and repeatability studies, together with absolute-drag and part-removal comparisons against GTU wind-tunnel measurements, with calibration choices and comparison limits documented (Appendix~\ref{app:vv}).
  \item \textbf{Benchmark.} Geometry-grouped splits with fixed metrics and reference results for \SMART{}, \DoMINO{}, \GeoT{} and \ABUPT{} on surface and volume tracks, with drag integrated from the predicted fields (\S\ref{sec:benchmark}).
  \item \textbf{Distribution shifts.} Graded out-of-distribution sets along the three axes, \ifdefined\ArxivVersion\else released with the dataset, \fi with the four models evaluated on each (\S\ref{sec:ood}).
\end{itemize}

%% file: sections/03-dataset.tex
\section{The \SHIFTTruck{} dataset}
\label{sec:dataset}

\subsection{Baseline geometry and design space}
\label{sec:dataset:design}

\begin{figure}[t]
\centering
\includegraphics[width=\linewidth]{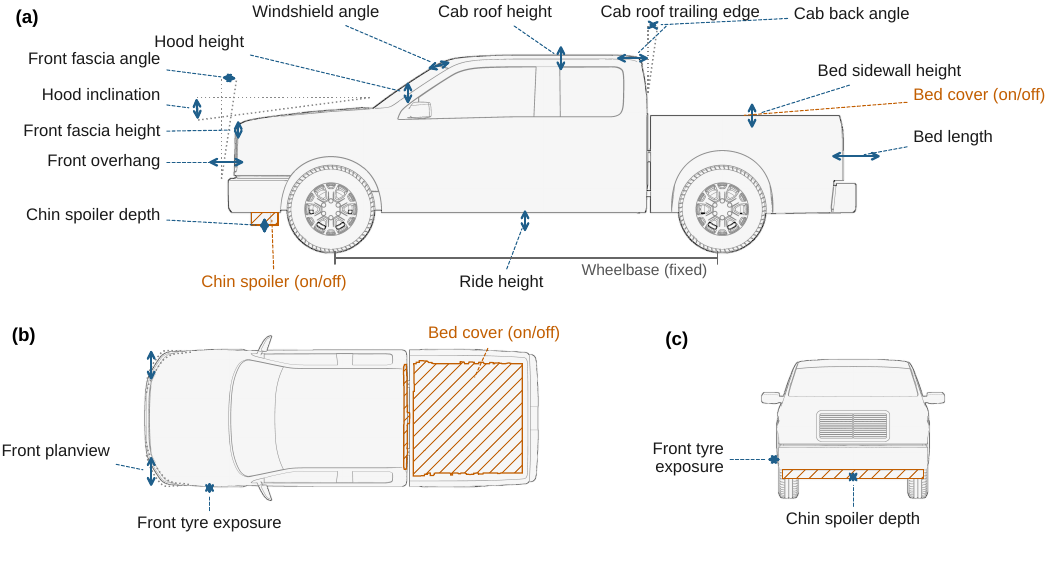}
\caption{Design parameters of \SHIFTTruck{} on the GTU baseline. (a) Side, (b) top and (c) front views. Blue arrows mark translational morphs. Dotted line pairs with an arrow across the open end mark angle morphs. Dotted outlines in (b) show the nose at the two extremes of front planview. Hatched regions are the two on/off parts. Ranges are given in Appendix~\ref{app:geometry}.}
\label{fig:param}
\end{figure}

The baseline geometry for SHIFT-truck is the Generic Truck Utility (GTU), the open reference geometry for pickup and SUV aerodynamics \citep{woodiga2020gtu,howard2021gtu}. We use the $4\times2$ pickup variant (Fig.~\ref{fig:param}), with a long cab, a short open bed, exposed mirrors and an airdam under the front fascia. The bed is a cavity between the cab and the tailgate, and the dataset carries both open-bed and closed-bed cases. With the bed open, the flow separates from the roof's trailing edge and the rear of the cab, a low-momentum recirculation fills the bed, and a second separation at the tailgate feeds the base wake (Fig.~\ref{fig:baseflow}c). A flush cover removes the cavity and leaves a continuous upper surface from the cab to the tailgate. The underbody is open in both states, with exposed wheels and drivetrain. \citet{woodiga2020gtu} measured wind-tunnel forces for the baseline and for single-part removals, which Appendix~\ref{app:vv} validates against.

Every case is built from this geometry. Fifteen continuous parameters move the front end, the cab, the bed and the ride height (Fig.~\ref{fig:param}). The body is bound to a deformation cage, so these morphs keep the surface topology and no case needs new CAD. A flush bed cover and a chin spoiler are switched on and off, which adds or removes a part and so changes the topology of the wetted surface. The wheels, underbody and drivetrain are not morphed, the wheelbase is fixed, and ride height moves the body relative to the wheels. Body length varies by about a meter across the dataset and the overall height by 0.4~m. Appendix~\ref{app:geometry} gives the cage, the parameter ranges and the dataset composition.

\subsection{Simulation}
\label{sec:dataset:sim}

\begin{figure}[t]
\centering
\includegraphics[width=\linewidth]{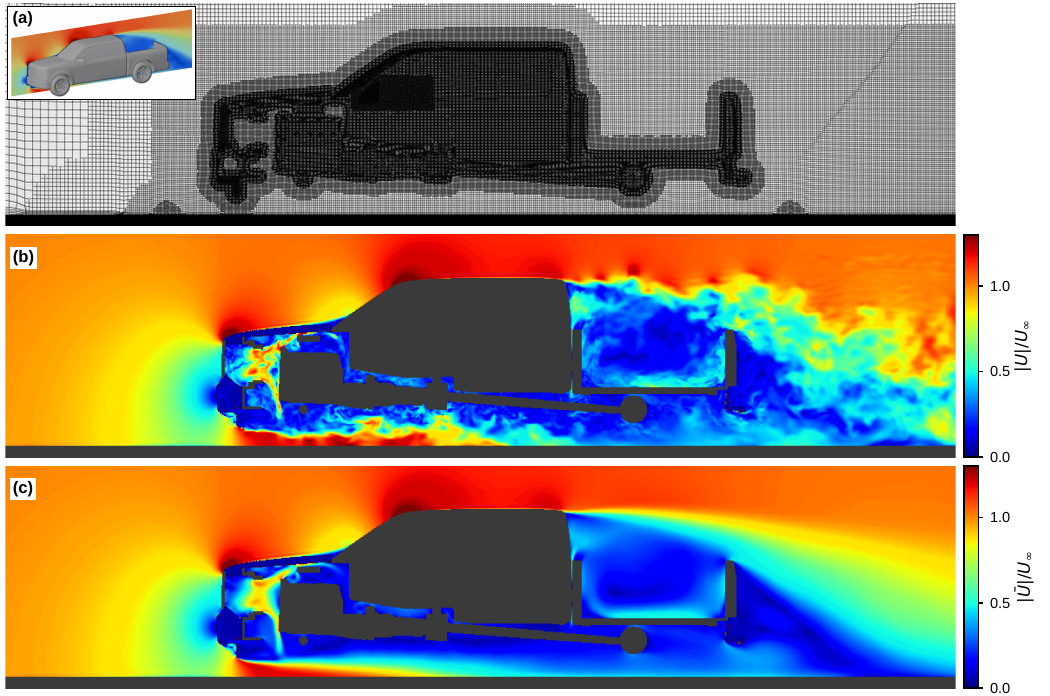}
\caption{The baseline GTU on the centreplane $y=0$, the vertical plane through the middle of the truck from nose to tail (inset). (a) Mesh. (b) Instantaneous velocity magnitude. (c) Time-averaged velocity magnitude.}
\label{fig:baseflow}
\end{figure}

We draw \nBaseTrucksPlanned{} base trucks from a Sobol sequence over the fifteen continuous parameters and simulate each with the four combinations of bed cover and chin spoiler, \ifdefined\ArxivVersion which gives \nSobolPoints{} cases, all of which are released.\else which gives \nSobolPoints{} planned cases, of which the release holds \nRelease{}.\fi{} Every case uses the same numerical setup, fixed by the study of Appendix~\ref{app:vv}. Turbulence is treated by DDES on the Spalart-Allmaras model \citep{spalart2006ddes}, with a Vreman subgrid model \citep{vreman2004eddy}, so the separations behind the cab and the tailgate are resolved as unsteady turbulence (Fig.~\ref{fig:baseflow}b) and the attached boundary layers stay modelled. Meshes contain about \nCellsCorpus{} million cells (Fig.~\ref{fig:baseflow}a), informed by the grid-refinement study of Appendix~\ref{app:vv}. Each case starts from a converged RANS solution, is time-averaged over the final \nAvgCTUCorpus{} convective time units, and takes \nWallClockCorpus{} on \nGPUsPerCase{} H100 GPUs. Appendix~\ref{app:cfd} gives the solver, the discretization, the mesh construction and the domain. Repeated simulations of the same case agree in drag to within \nRepeatCDHi{}\%. On the reference GTU baseline, simulated \CD{} exceeds the wind-tunnel measurement of \citet{woodiga2020gtu} by \nAbsOffsetCD{}. The mirror- and airdam-removal drag changes used during setup calibration differ from measurements by at most \nValTol{}. Appendix~\ref{app:vv} reports the grid study, subsequent changes to the dataset setup, repeatability and these experimental comparisons.

\subsection{Dataset Contents}
\label{sec:dataset:aero}

\begin{figure}[t]
\centering
\includegraphics[width=\linewidth]{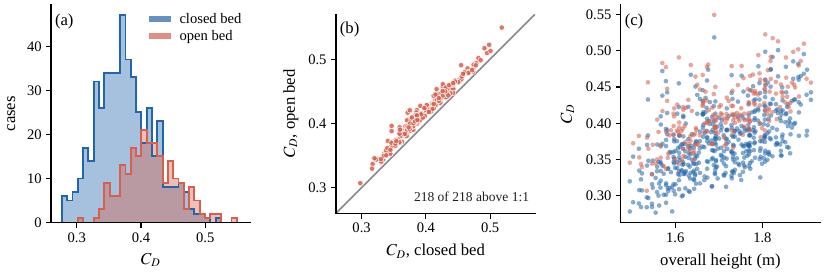}
\caption{The design space of \SHIFTTruck{}. (a) Drag distribution by bed state over the \nCorpus{} cases of the benchmark set. (b) The \nBedPairsMatched{} base trucks carrying both bed states with every other parameter held fixed, each plotted as its closed-bed against its open-bed drag, with the 1:1 line. (c) Drag against overall height, coloured by bed state.}
\label{fig:designspace}
\end{figure}

Fig.~\ref{fig:designspace}a shows the drag distribution over the dataset. It spans roughly a factor of two, and the closed-bed and open-bed populations occupy different parts of that range with a broad overlap, and overall height is the strongest single driver of the spread (Fig.~\ref{fig:designspace}c, Appendix~\ref{app:aero}). Designing the bed is crucial, since bed length, sidewall height, ride height and the cover are all chosen for utility, and each affects the drag. Published pickup aerodynamics studies a handful of configurations at a time \citep{algarni2010pickup,cfd_pickup_drag,woodiga2020gtu}, and no open dataset maps these trade-offs across a design space. Because drag force scales with frontal area, each drag count costs more fuel or range on a pickup than on a passenger car.

For machine learning, the dataset poses a harder problem than a single closed body. The target responds to a part appearing and disappearing from the wetted surface, which a continuous morph does not produce. \nBedPairsMatched{} base trucks carry both bed states with every other parameter held fixed, and opening the bed raises drag on every one of them (Fig.~\ref{fig:designspace}b), so a surrogate can be scored on a controlled comparison as well as on held-out shapes. The chin spoiler gives a second such comparison on the closed-bed trucks.

% \subsection{Contents and access}
% \label{sec:dataset:contents}

Each case is released as a surface in STL and VTK PolyData, carrying time-averaged pressure and wall shear stress, and a volume in VTK UnstructuredGrid, carrying time-averaged pressure and velocity, together with its force coefficients, reference values and a metadata record. The dataset is released under the CC-BY-NC-4.0 license\ifdefined\ArxivVersion\else, with its Croissant metadata \citep{akhtar2024croissant}\fi{} and is available \ifdefined\ArxivVersion at \ShiftTruckURL\else for review at \texttt{gs://shift-truck-sample/}\fi.

%% file: sections/04-benchmark.tex
\section{Benchmark}
\label{sec:benchmark}

\begin{table}[t]
\caption{Benchmark results on the \nTest{} test cases. Error metrics are means over cases and seeds 42, 43 and 44. Drag $R^2$ entries give the mean $\pm$ the standard deviation across seeds. Params is the trainable parameter count in millions and GPU-h the training cost in H100 hours for \nTrainEpochs{} epochs in FP32. Inf.\ is the wall-clock time to predict one case on a surface of \nInfFaces{} million faces, on one NVIDIA T4 GPU in float32.}
\label{tab:results}
\centering
\small
\setlength{\tabcolsep}{4pt}
\resizebox{\linewidth}{!}{\input{tables/t_results_body}}
\end{table}

\textbf{Models and tracks.}
We benchmark four neural surrogates that map a raw geometry to field values at arbitrary query points, \DoMINO{} \citep{domino2025}, \GeoT{} \citep{geotransolver2025}, \ABUPT{} \citep{abupt2025} and \SMART{} \citep{smart2026}. Each is trained for \nTrainEpochs{} epochs in FP32 on a surface track (pressure and wall shear stress on the body) and a volume track (pressure and velocity in the fluid), using its reference architecture and optimizer. \GeoT{} and \DoMINO{} minimize mean squared error, \ABUPT{} uses an $L_1$ loss, and \SMART{} uses relative $L_2$ loss. Appendix~\ref{app:benchmark} gives the architectures and training configurations.

\textbf{Split.}
The benchmark uses the \nCorpus{} cases that had completed when the split was frozen. We split by base truck, so that no geometry appears in two sets under any bed or spoiler setting. Validation and test trucks are drawn from the \nBedPairs{} trucks that carry both bed states, each contributing its two closed-bed cases and its open-bed case. They are selected deterministically on a quantile grid over overall height and body length, the two main morph axes, taking in each cell the truck nearest its centre, on a $6\times5$ grid for test and a $6\times4$ grid for validation. Test trucks lie as close to the training set in parameter space as training trucks lie to each other (Appendix~\ref{app:split}). The split has \nTrain{} training, \nVal{} validation and \nTest{} test cases, and each of the \nTestTrucks{} test trucks appears closed with and without the chin spoiler, and open.

\textbf{Metrics.}
Fields are scored in non-dimensional form on the solver's own surface polygons and volume cells. All four models use the full volume mesh for evaluation. The pressure coefficient is scored by its mean absolute error, and wall shear stress and velocity by their relative $L_2$ error. Drag is integrated from the predicted surface fields and no model is trained on forces.

\textbf{Results.}
Table~\ref{tab:results} gives the results on the test set together with the variation of drag $R^2$ across training seeds. On the surface, \GeoT{} has the lowest error on both fields, with a pressure error of about \nSgeoCpPct{}\% of the dynamic pressure, and \SMART{} is within 2\% of it on shear stress. In the volume, \SMART{} has the lowest velocity error at about half the training cost of \GeoT{} and \DoMINO{}, and the three leading models are within 10\% of each other on pressure. A ranking on the body therefore does not carry over to the wake. \SMART{}, \GeoT{} and \ABUPT{} recover drag with $R^2$ between \nCDRsqLeadLo{} and 0.97. Across three training seeds, the standard deviation of drag $R^2$ reaches \nSeedSdSMART{}, which is comparable to the differences between the three. The lowest drag error, from \SMART{}, is \nSsmartCDMAE{} in \CD{}, about \nSsmartDragN{}\,N at \nUinfKmh{}\,km/h. \DoMINO{} trails on every channel, and its drag $R^2$ falls to 0.03. 

\begin{figure}[t]
\centering
\includegraphics[width=\linewidth]{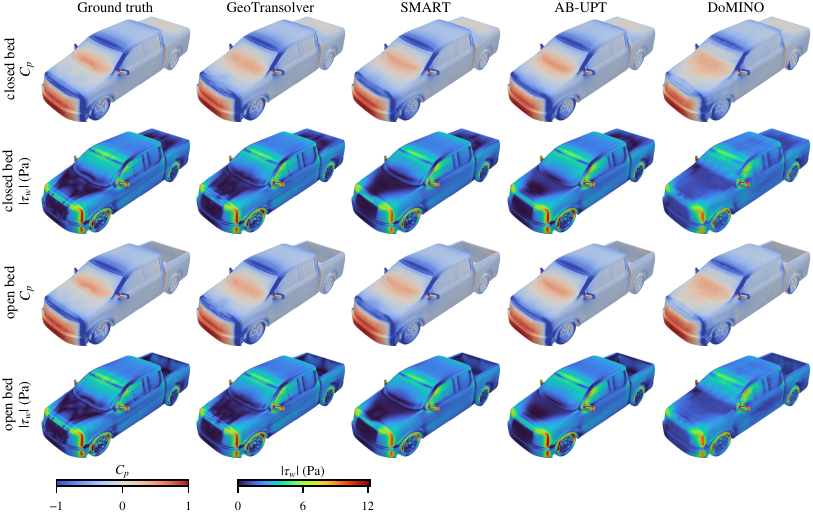}
\caption{Surface pressure coefficient and wall-shear-stress magnitude for a matched pair of test trucks, the same base geometry with the bed closed (top two rows) and open (bottom two rows). Columns are ground truth and the four surface-track models.}
\label{fig:fields}
\end{figure}

\begin{figure}[t]
\centering
\includegraphics[width=\linewidth]{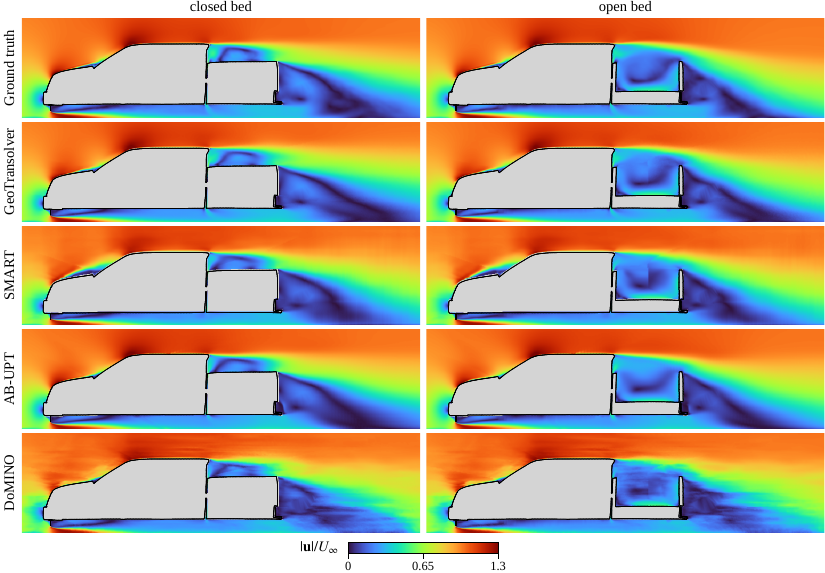}
\caption{Time-averaged velocity magnitude on the centerplane $y=0$ for the pair of Fig.~\ref{fig:fields}, with the bed closed (left) and open (right). Rows are ground truth and the four volume-track models.}
\label{fig:planes}
\end{figure}

\textbf{Surface.}
Figure~\ref{fig:fields} shows the surface fields of a matched pair (errors in Appendix~\ref{app:surface}). The flow stagnates on the grille and fascia, decelerates into the corner at the base of the windshield, and accelerates around the A-pillar and the mirrors. This pressure field is set by the outer flow, which the body shape largely determines, and all four models reproduce it, together with the high wall shear stress along the hood and roof leading edges, the A-pillar and the wheel-arch lips. The models differ where the shear stress depends on the state of the boundary layer. Between the hood leading edge and the windshield the near-wall flow slows under an adverse pressure gradient, and the true field shows a region of low shear on the hood, broken into streamwise streaks. \GeoT{} reproduces the region and its streaks most closely, and \SMART{} and \ABUPT{} smooth out the streaks. \DoMINO{} fills in every low-shear region: its prediction shows no deceleration ahead of the windshield, no thickening of the boundary layer toward the roof trailing edge, and no slow recirculating flow on the bed floor. Low shear marks where the near-wall flow decelerates, separates or recirculates, and on a bluff body these regions set the pressure drag. \DoMINO{}'s pressure also carries a regional offset over the hood and grille, which the drag, a small difference between large forces on the front and rear of the body, inherits.

\textbf{Volume.}
Figure~\ref{fig:planes} shows the centreplane velocity for the same pair (errors and pressure in Appendix~\ref{app:volume}). Every model reproduces the acceleration over the hood and roof, which the body shape sets through the outer flow. Every model also separates the flow at the roof trailing edge and, on the open truck, fills the bed with a recirculation. The geometry fixes both, since the flow leaves a sharp edge at a fixed point and a cavity of this depth traps a recirculation whatever the state of the incoming boundary layer. The largest errors sit in the shear layer leaving the cab roof and in the near wake behind the tailgate, the regions of turbulent mixing that the geometry does not fix. The spreading of the shear layer sets how much momentum it exchanges with the bed and the wake, and the length of the near wake is tied to the base pressure and so to the pressure drag. Every model predicts the velocity less accurately with the bed open, and the added error sits in the bed recirculation and the shear layer above it, which exchange momentum. \SMART{}, \GeoT{} and \DoMINO{} also show a step in velocity halfway along the bed with no counterpart in the true flow, possibly due to a signed distance function as model input. \DoMINO{} recovers the gross wake, but its field is noisy in the outer flow, where the true field is smooth, and its bed recirculation is blurred, matching its filled-in shear stress on the bed floor.

%% file: tables/t_results_body.tex
\begin{tabular}{lrrrcccc@{\hskip 1.4em}rrcc}
\toprule
 & \multicolumn{7}{c}{Surface track} & \multicolumn{4}{c}{Volume track} \\
\cmidrule(lr){2-8}\cmidrule(lr){9-12}
Model & Params & GPU-h & Inf.\ (s) & $C_p$ MAE & $\WSS$ \relLtwo{} & \CD{} MAE & \CD{} $R^2$ & Params & GPU-h & $C_p$ MAE & $\bm u$ \relLtwo{} \\
\midrule
\GeoT{}   & 21.7 & 91 & \nInfGeoT{} & 0.021 & 0.173 & 0.0069 & $0.958 \pm \nSeedSdGeoT{}$ & 22.2 & 144 & 0.029 & 0.149 \\
\SMART{}  & 12.6 & 61 & \nInfSMART{} & 0.023 & 0.175 & 0.0047 & $0.971 \pm \nSeedSdSMART{}$ & 12.8 & 69 & 0.028 & 0.114 \\
\ABUPT{}  & 16.4 & 54 & \nInfABUPT{} & 0.025 & 0.210 & 0.0077 & $0.953 \pm \nSeedSdABUPT{}$ & 41.3 & 89 & 0.027 & 0.132 \\
\DoMINO{} & 10.1 & 68 & \nInfDoMINO{} & 0.051 & 0.605 & 0.0345 & $0.034 \pm \nSeedSdDoMINO{}$ & 18.5 & 132 & 0.046 & 0.163 \\
\bottomrule
\end{tabular}

%% file: sections/06-ood.tex
\section{Out-of-distribution evaluation}
\label{sec:ood}
Distribution shifts in Scientific Machine Learning arise not only from changes in the physical problem, such as the geometry or the operating conditions, but also from changes in its numerical representation, as the same field may be represented on meshes of different resolution or connectivity. This discretization shift is particularly important for neural operators, whose intended advantage is to approximate mappings between function spaces rather than mappings tied to a particular discretization.

We evaluate the four trained models of \S\ref{sec:benchmark} along each axis. For the operating point, a training-set truck is re-simulated at 100, 110, 120 and 130\,km/h on the same mesh and with the same solver settings, while all training data are at \nUinfKmh{}\,km/h. For the discretization, the input surface of the same truck at \nUinfKmh{}\,km/h is decimated by quadric-error-metric edge collapse \citep{garland1997qem} to 50, 30, 20 and 10\% of its \nInfFaces{} million faces, and the prediction is compared with the solver drag of the full-resolution case. At 10\% of the faces the decimated surface lies within 7.6\,mm of the original, and integrating the solver fields over it changes \CD{} by 1 count. The shift therefore changes how the geometry is represented but leaves the geometry and the flow essentially unchanged. For the vehicle archetype, the truck-trained models predict ten SHIFT-SUV cases \citep{shift_placeholder}, and a model trained on SHIFT-SUV gives the in-distribution reference. Every error is the absolute drag-coefficient error in counts, where one count is $10^{-3}$ in \CD{}. These OOD sets are released with the dataset. 
%The truck on the first two axes is a training case, so its error at the training condition is a training-set error.
%\note{State which checkpoint each model uses here (surface track or combined), and whether the four sweep cases are released.}

\begin{figure}[t]
\centering
\includegraphics[width=\linewidth]{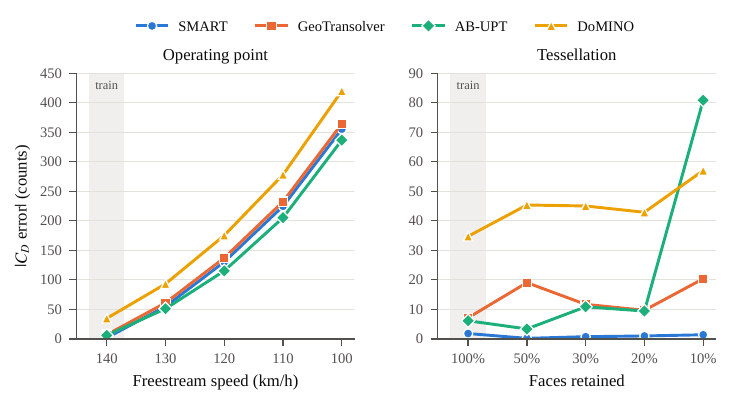}
\caption{Absolute drag error of the four truck-trained models against freestream speed (left) and the fraction of input surface faces retained (right). The shaded column is the training condition, a training-set truck at \nUinfKmh{}\,km/h on its full surface. One count is $10^{-3}$ in \CD{}.}
\label{fig:ood}
\end{figure}

\textbf{Operating point.}
All four models receive freestream speed as an input, but it is constant during training. Without physical rescaling, drag error grows with the speed offset and reaches 337 to 420 counts at 100\,km/h (Fig.~\ref{fig:ood}, left), while the solver \CD{} changes by less than 7 counts. Predicted drag force is unchanged for \SMART{}, \GeoT{} and \DoMINO{}, and varies by less than 5\% for \ABUPT{}. At fixed reference density and area, constant predicted force gives $\widehat{C}_D(U)=\widehat{C}_D(U_{\mathrm{train}})(U_{\mathrm{train}}/U)^2$, which accounts for nearly all of the error growth. Multiplying a speed-invariant force prediction by $(U/U_{\mathrm{train}})^2$ would instead keep its predicted \CD{} at the training-speed value; the remaining variation in its error would follow the solver's \CD{} variation. This sweep therefore diagnoses missing dynamic-pressure scaling in the uncorrected predictions of these single-speed-trained models and motivates the speed correction below.

\textbf{Discretization.}
The error of \SMART{} stays below 2 counts at every decimation level (Fig.~\ref{fig:ood}, right). \GeoT{} lies between 9.7 and 20.4 counts, above its full-surface error of 7 counts at every level. \ABUPT{} stays within 11 counts down to 20\% of the faces and reaches 81 counts at 10\%. \DoMINO{} rises from 35 counts on the full surface to 57 counts at 10\%.

\textbf{Vehicle archetype.}
The SUV cases are at 31.3\,m/s (112.7\,km/h), so the truck-trained predictions are rescaled by $(U/U_{\mathrm{train}})^2$ before scoring. After rescaling, every truck-trained model underpredicts \CD{} on all ten SUVs (Table~\ref{tab:ood_geometry}). Mean errors range from 67 counts for \GeoT{} to 251 counts for \SMART{}. The same models are within 2 to 35 counts on the training-set truck. The model trained on SHIFT-SUV predicts the ten cases with a mean error of 1.5 counts. Without rescaling, the speed offset adds an overprediction that partly cancels the geometry error, and the raw column of Table~\ref{tab:ood_geometry} carries both effects.

\input{tables/t_ood_geometry}

% \paragraph{Detecting the shift.}
% \note{TODO. Intro bullet 4 promises shift detection. Data exist for \SMART{} only. Its detector rejects all five SUVs and reports in-distribution with confidence 1.0 at every speed and every decimation level, including 100\,km/h, where the drag error is 355 counts.}

%% file: tables/t_ood_geometry.tex
\begin{table}[t]
\centering
\caption{Absolute drag error on ten SHIFT-SUV cases in counts ($10^{-3}$ in \CD{}), averaged over the cases. Truck-trained predictions are rescaled by $(U/U_{\mathrm{train}})^2$ to the SUV speed in the rescaled column and left as predicted in the raw column. The truck column is each model's error on the training-set truck at \nUinfKmh{}\,km/h. The reference model is trained on SHIFT-SUV.}
\label{tab:ood_geometry}
\begin{tabular}{lccc}
\toprule
Model & Truck & SUV, rescaled & SUV, raw \\
\midrule
\SMART{} & 1.8 & 251 & 214 \\
\GeoT{} & 7.1 & 67 & 81 \\
\ABUPT{} & 6.1 & 241 & 199 \\
\DoMINO{} & 34.8 & 147 & 54 \\
\midrule
SHIFT-SUV reference & -- & 1.5 & 1.5 \\
\bottomrule
\end{tabular}
\end{table}

%% file: sections/07-conclusion.tex
\section{Conclusion}
\label{sec:conclusion}

Pickup trucks are among the highest-selling and yet, least aerodynamic vehicles in the United States and the move to electric powertrains turns their drag into a question of battery size and range. We introduced \SHIFTTruck{}, to our knowledge the first open, high-fidelity aerodynamics dataset for this vehicle class, with \nRelease{} SA-DDES simulations of a parametric pickup built on the GTU, each on a mesh of about \nCellsCorpus{} million cells. The setup is verified on a \nGridLevels{}-level grid and reproduces measured drag changes on the same truck to within \nValTol{}. Generating the dataset took about \nGPUhCorpus{} H100 GPU-hours.

In our benchmark of four recent neural surrogates, every model reproduces the stagnation on the front, the separation at the roof trailing edge and the recirculation in the open bed, all of which are dictated by the body shape. The errors concentrate in the shear layer leaving the cab roof and in the near wake, the regions of turbulent mixing that set the base pressure and therefore the pressure drag. Every model is less accurate with the bed open and a ranking on the surface does not carry over to the volume. Under the controlled distribution shifts, no model responds to a change in freestream speed, the sensitivity to surface decimation differs by more than an order of magnitude between models and every model underpredicts drag on SUVs. The open-bed recirculation and the parts that change the wetted surface make \SHIFTTruck{} a demanding test for surrogate architectures and for distribution shift in physics-based machine learning, as well as a basis for data-driven design of this vehicle class.

\section{Limitations and future work}
\label{sec:limitations}

\SHIFTTruck{} describes the time-averaged flow, as a result of which the data cannot train models of unsteady loads or of the vortex shedding behind the tailgate. Every geometry is morphed from one GTU configuration, so the dataset spans variation within a single vehicle archetype and not the range of production pickup trucks. The benchmark scores surrogates against the solver, and its targets carry the modelling assumptions of DDES with wall functions. Training uses one freestream speed at zero yaw. The speed sweep of \S\ref{sec:ood} checks deployment of these fixed models on one training geometry, where the solver drag coefficient varies only weakly with speed. Crosswind remains outside the dataset. The experimental comparison covers the GTU baseline and two part removals used during setup calibration, with a baseline drag offset of \nAbsOffsetCD{} against measurement (Appendix~\ref{app:vv}). The benchmark evaluates point predictions, and seed variation is reported for drag only.

We continue to add steady-state cases to the current design space. Future work will extend \SHIFTTruck{} with transient simulations, so that unsteady loads and wake dynamics can be learned alongside the mean flow. The errors of current surrogates in the shear layer and the near wake call for architectures that represent turbulent mixing more faithfully. Uncertainty quantification for field surrogates and their use in data-driven design optimization of pickup trucks are further natural directions.

%% file: appendix/A-geometry.tex
\section{Geometry and parameterization}
\label{app:geometry}

\subsection{Baseline}
The Generic Truck Utility (GTU) is a family of open reference geometries for pickups and SUVs \citep{woodiga2020gtu,howard2021gtu}, distributed by the European Car Aerodynamic Research Association (ECARA). \citet{howard2024egtu} describe an electrified variant. We morph configuration~7 of the first release, a $4\times2$ pickup with a long cab, a short bed and a wheelbase of 3.27~m. It has an open cargo bed closed by a tailgate, exposed mirrors, an airdam below the front fascia, exposed wheels and a detailed underbody with drivetrain. \citet{woodiga2020gtu} measured the forces on this configuration in a wind tunnel, with and without single parts, and simulated the same configurations. Appendix~\ref{app:vv} validates our setup against both.

\subsection{Deformation cage}
The body is bound to a deformation cage of \nCageVerts{} vertices with the harmonic-coordinate mesh-deform binding of Blender. The cage carries the cab, the bed, the hood, the underhood, the bed cover and the mirrors. The wheels, the underbody and the drivetrain sit outside the cage and keep their baseline shape. The wheelbase is fixed as a result, and the ride-height parameter moves the body relative to the wheels. A second cage of eight vertices carries the chin spoiler. Its depth is an independent parameter. Three further shape keys on this cage follow front overhang, front tyre exposure and ride height, and they keep the spoiler attached to the fascia as the body morphs.

\subsection{Parameters}
Table~\ref{tab:params} lists the \nParamsCont{} continuous parameters. Each is a shape key with its own slider range, and a slider value of zero gives the baseline geometry. The ranges are asymmetric about the baseline. Nine keys run from $-0.5$ to $+1.0$ and four from $-0.5$ to $+0.5$. Front overhang runs from $-0.5$ to $+0.75$ and bed sidewall height from $-1.0$ to $+1.0$. The sampler draws a normalized value in $[-1,1]$ for each parameter and maps it linearly onto the slider range. In normalized coordinates the baseline sits at $-0.33$ for the nine keys that extend to $+1.0$, at $-0.20$ for front overhang and at zero for the five symmetric keys. The centre of the design space is a truck with a longer bed, a higher roof and a more swept nose than the baseline. Front planview moves the outboard corners of the nose inboard with the centreline held fixed, which changes the plan-view taper of the front end.

\subsection{Discrete parts}
The bed cover is a flush tonneau panel over the cargo bed. With the cover in place the bed is a closed box. Without it the bed is an open cavity between the cab back, the sidewalls and the tailgate. The chin spoiler extends the lower fascia downward and restricts the flow into the underbody. Both parts are added to or removed from the simulated assembly, and each change alters the topology of the wetted surface. The chin spoiler is a separate part from the GTU airdam removed in the validation of Appendix~\ref{app:vv}.

\subsection{Sampling}
The continuous parameters are sampled at \nBaseTrucksPlanned{} points of a scrambled Sobol sequence \citep{sobol1967distribution}, generated with the \texttt{qmc.Sobol} sampler of SciPy at seed 42. Each point defines a base truck. Every base truck is simulated with all four combinations of bed cover and chin spoiler, for \nSobolPoints{}\ifdefined\ArxivVersion\else{} planned\fi{} cases. Case ids 0000--0249 carry both parts, 0250--0499 the cover alone, 0500--0749 the spoiler alone and 0750--0999 neither. The ids $i$, $i+250$, $i+500$ and $i+750$ belong to the same base truck. Each case is meshed and simulated independently. The benchmark of \S\ref{sec:benchmark} uses the \nCorpus{} cases that had completed at the training freeze, and the public release holds \ifdefined\ArxivVersion all \nRelease{}\else\nRelease{}\fi{} cases. Table~\ref{tab:composition} gives the composition of both sets.

\input{tables/t_parameters}

\begin{table}[h]
\caption{Composition of the benchmark set, the \nCorpus{} cases complete at the training freeze, and of the release.}
\label{tab:composition}
\centering
\small
\begin{tabular}{lr}
\toprule
Cases at the freeze & \nCorpus{} \\
\quad closed bed / open bed & \nClosed{} / \nOpen{} \\
Distinct base trucks & \nBaseTrucks{} \\
\quad appearing three, two, one times & 218 / 28 / 2 \\
\quad carrying both bed states & \nBedPairs{} \\
\quad carrying both bed states, chin spoiler held fixed & \nBedPairsMatched{} \\
\quad carrying both spoiler states, bed closed & \nSpoilerPairs{} \\
Body length (m) & \nBodyLenLo{} to \nBodyLenHi{} \\
Overall height (m) & \nHeightLo{} to \nHeightHi{} \\
Wetted area (m$^2$) & \nWettedLo{} to \nWettedHi{} \\
Cases in the public release & \nRelease{} \\
\ifdefined\ArxivVersion\else
Cases planned & \nSobolPoints{} \\
\fi
\bottomrule
\end{tabular}
\end{table}

%% file: tables/t_parameters.tex
% Converted from ../v3/tables/t02_parameters.tex (GENERATED from
% config_7_parameterization.blend shape keys by viz_rough/f01_pipeline).
% Regenerate there and re-convert rather than editing values by hand.
\begin{table}[h]
\caption{The fifteen continuous design parameters. Each is a shape key on the deformation cage. The sampler draws a normalized value in $[-1,1]$ and maps it linearly onto the slider range. Displacement is the largest cage-vertex displacement at each end of the range, and the last column gives the normalized value of the baseline geometry.}
\label{tab:params}
\centering
\small
\resizebox{\linewidth}{!}{%
\begin{tabular}{llccc}
\toprule
Parameter & Direction & Slider range & Displacement (m) & Baseline \\
\midrule
Front overhang & $x$, forward & $-0.50$ to $+0.75$ & $-0.154$ to $+0.231$ & $-0.20$ \\
Hood height & $z$ & $-0.5$ to $+1.0$ & $-0.046$ to $+0.092$ & $-0.33$ \\
Hood inclination & $z$ at hood front (rotation about the cowl) & $-0.5$ to $+0.5$ & $-0.192$ to $+0.192$ & $+0.00$ \\
Front fascia height & $z$ & $-0.5$ to $+0.5$ & $-0.125$ to $+0.125$ & $+0.00$ \\
Front fascia angle & $x$ at fascia top (rotation about its base) & $-0.5$ to $+1.0$ & $-0.188$ to $+0.375$ & $-0.33$ \\
Front planview & $y$, corners inboard & $-0.5$ to $+1.0$ & $-0.225$ to $+0.451$ & $-0.33$ \\
Front tyre exposure & $y$, outboard & $-0.5$ to $+0.5$ & $-0.050$ to $+0.050$ & $+0.00$ \\
Windshield angle & $x$ at windshield top & $-0.5$ to $+1.0$ & $-0.098$ to $+0.196$ & $-0.33$ \\
Cab roof height & $z$ & $-0.5$ to $+1.0$ & $-0.111$ to $+0.222$ & $-0.33$ \\
Cab roof trailing edge & $x$, rearward & $-0.5$ to $+1.0$ & $-0.152$ to $+0.305$ & $-0.33$ \\
Cab back angle & $x$ at cab-back top & $-0.5$ to $+1.0$ & $-0.114$ to $+0.227$ & $-0.33$ \\
Bed length & $x$, rearward & $-0.5$ to $+1.0$ & $-0.242$ to $+0.484$ & $-0.33$ \\
Bed sidewall height & $z$ & $-1.0$ to $+1.0$ & $-0.139$ to $+0.139$ & $+0.00$ \\
Ride height & $z$, body relative to wheels (down) & $-0.5$ to $+0.5$ & $-0.061$ to $+0.061$ & $+0.00$ \\
Chin-spoiler depth & $z$, downward & $-0.5$ to $+1.0$ & $-0.048$ to $+0.096$ & $-0.33$ \\
\midrule
Bed cover & on/off & --- & --- & --- \\
Chin spoiler & on/off & --- & --- & --- \\
\bottomrule
\end{tabular}}
\end{table}

%% file: appendix/B-cfd.tex
\section{Numerical setup}
\label{app:cfd}

\subsection{Solver}
The flow is solved with a GPU-native, cell-centred finite-volume code on unstructured polyhedral meshes \citep{krakos2025gpu,economon2024autocfd4}. The gas is ideal, and the viscosity follows Sutherland's law. The convective flux uses a hybrid upwind--centred scheme with low dissipation in the resolved regions of the flow and stable behaviour in the coarse far field. Face states are reconstructed from cell-centred values and gradients to second order in space, without a limiter. Time integration is second-order implicit (BDF2) with five inner iterations per step. Every operation runs on the GPU in double precision.

\subsection{Turbulence}
Turbulence is treated by delayed detached-eddy simulation on the Spalart--Allmaras model \citep{spalart2006ddes}, with the Vreman subgrid model \citep{vreman2004eddy}. The delayed formulation shields the attached boundary layer, which stays modelled. This prevents modelled-stress depletion and the premature separation it would cause.

\subsection{Time integration and averaging}
Each case starts from a converged RANS solution of \nRANSIters{} iterations. The time step is ramped down between iterations 1,000 and 1,500 of the transient run and then held at $\Delta t = \nDt$~s. A case runs \nSteps{} iterations, about \nPhysTime{}~s of physical time. Fields and forces are averaged over the final \nAvgStepsCorpus{} steps. This window lasts about \nAvgTimeCorpus{}~s, or \nAvgCTUCorpus{} convective time units of $L/\Uinf = \nCTU$~s. The validation runs of Appendix~\ref{app:vv} average over a longer window of \nAvgStepsVal{} steps.

\subsection{Mesh}
\begin{figure}[h]
\centering
\includegraphics[width=0.8\linewidth]{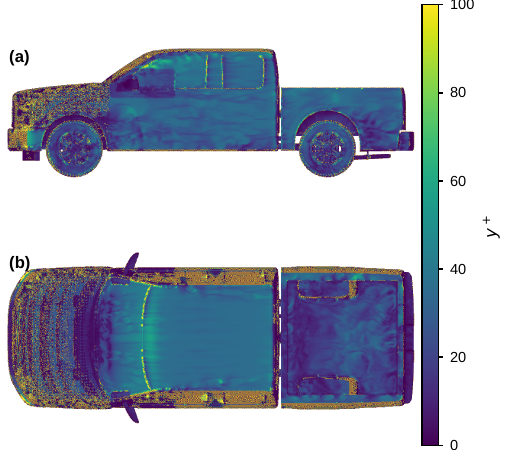}
\caption{Wall $\yplus$ on the baseline GTU with the dataset setup, (a) side and (b) underside. The body panels are wall-modelled at $\yplus$ of \nYplusLo{} to \nYplusHi{}. The mirrors, the chin spoiler and the wheel edges resolve below 5.}
\label{fig:yplus}
\end{figure}

Meshes are generated with snappyHexMesh \citep{weller1998openfoam} following the practice of the Automotive CFD Prediction Workshop \citep{hupertz2022autocfd2,economon2024autocfd4}. Refinement follows the body through distance-based shells around the surface, with additional refinement at feature edges. The surface refinement level is 8 on the body panels, 9 on the wheels, 10 on the chin spoiler and 11 on the mirrors. Dedicated boxes refine the mirror wakes, which distance-based sizing alone leaves under-resolved. The near-wall mesh has \nBLLayers{} prism layers at an expansion ratio of \nBLER{}, and every vehicle surface uses wall functions. The body panels sit at $\yplus$ of \nYplusLo{} to \nYplusHi{}. The mirrors and the chin spoiler carry the finest surface refinement and sit at \nYplusFineLo{} to \nYplusFineHi{} (Fig.~\ref{fig:yplus}). The baseline GTU mesh has \nCellsBase{} million cells. The morphed geometries have about \nCellsCorpus{} million, and \texttt{variant\_0000} has 97.6 million cells and 103.2 million points.

\subsection{Domain and boundary conditions}
The domain is a closed box \nDomX{}~m long, \nDomY{}~m wide and \nDomZ{}~m high, the size of the CFD domain of \citet{woodiga2020gtu}. The vehicle sits about 21~m downstream of the inlet and 32~m upstream of the outlet. The inlet is a far-field boundary and the outlet a pressure outlet. The ceiling, the side walls and the floor ahead of the vehicle are slip walls. Under and behind the vehicle the floor is no-slip. It contains \nBelts{} belts moving at the freestream speed, one under each wheel and a main belt between them, which reproduce the rolling road of the experiment. The wheels rotate at 105.76~rad\,s$^{-1}$, which matches the belt speed at their radius of 0.3677~m. The released surface excludes the tunnel boundaries. The vehicle surface ends where the tyres meet the belts, and the contact patches are open.

\subsection{Operating point and reference quantities}
Every case runs at a single operating point of 140~km\,h$^{-1}$ at zero yaw, with a freestream static temperature of 293.15~K (Table~\ref{tab:refquant}). The force coefficients and the pressure coefficient are normalized with constant reference values of the baseline geometry, which the release stores in \texttt{params.json}. These include a reference temperature of 288.15~K. The Reynolds number and the blockage ratio are derived from these values. \citet{woodiga2020gtu} state neither for the experiment. Each case runs on \nGPUsPerCase{} H100 GPUs in about \nWallClockCorpus{}. At the validation settings the baseline GTU takes \nWallClockVal{} on \nGPUsVal{} GPUs and \nWallClockScale{} on \nGPUsScale{}.

\begin{table}[h]
\caption{Operating point, reference quantities and domain, common to every case.}
\label{tab:refquant}
\centering
\small
\begin{tabular}{llr}
\toprule
Quantity & Symbol & Value \\
\midrule
Freestream speed & \Uinf{} & \nUinf{} m\,s$^{-1}$ \\
Freestream static temperature & $T_\infty$ & \nTstatic{} K \\
Freestream density (derived) & $\rho_\infty$ & \nRhoFree{} kg\,m$^{-3}$ \\
Reynolds number $\Uinf L/\nu$ (derived) & \Rey{} & $\nRe$ \\
\midrule
Reference length (body length) & $L$ & \nRefLen{} m \\
Reference area (frontal) & \Aref{} & \nRefArea{} m$^2$ \\
Reference pressure & \pref{} & \nPref{} Pa \\
Reference temperature & $T_{\mathrm{ref}}$ & \nTref{} K \\
Reference density (derived) & $\rho_{\mathrm{ref}}$ & \nRho{} kg\,m$^{-3}$ \\
\midrule
Convective time unit & $L/\Uinf$ & \nCTU{} s \\
Time step & $\Delta t$ & $\nDt$ s \\
Domain, streamwise $\times$ spanwise $\times$ vertical & --- & $\nDomX \times \nDomY \times \nDomZ$ m \\
Blockage ratio (derived) & --- & \nBlockage{}\% \\
\bottomrule
\end{tabular}
\end{table}

%% file: appendix/C-verification-validation.tex
\section{Verification and validation}
\label{app:vv}

\begin{table}[h]
\caption{Validation against the wind-tunnel experiment and the CFD of \citet{woodiga2020gtu} on the GTU baseline and two single-part removals. Deltas are referenced to each source's own baseline. The reference CFD is reported as deltas only.}
\label{tab:validation}
\centering
\small
\begin{tabular}{lcc@{\hskip 1.5em}ccc}
\toprule
 & \multicolumn{2}{c}{Absolute \CD{}} & \multicolumn{3}{c}{\dCD{} from baseline} \\
\cmidrule(lr){2-3}\cmidrule(lr){4-6}
Configuration & Ours & Experiment & Ours & Experiment & Reference CFD \\
\midrule
$4\times2$ baseline & \nBaseCD{} & \nExpBaseCD{} & 0 & 0 & 0 \\
No mirrors & 0.435 & 0.381 & $\nDeltaMirrorLum$ & $\nDeltaMirrorExp$ & $\nDeltaMirrorRef$ \\
No airdam  & 0.481 & 0.421 & $\nDeltaAirdamLum$ & $\nDeltaAirdamExp$ & $\nDeltaAirdamRef$ \\
\bottomrule
\end{tabular}
\end{table}

\subsection{Grid refinement}
\begin{figure}[h]
\centering
\includegraphics[width=\linewidth]{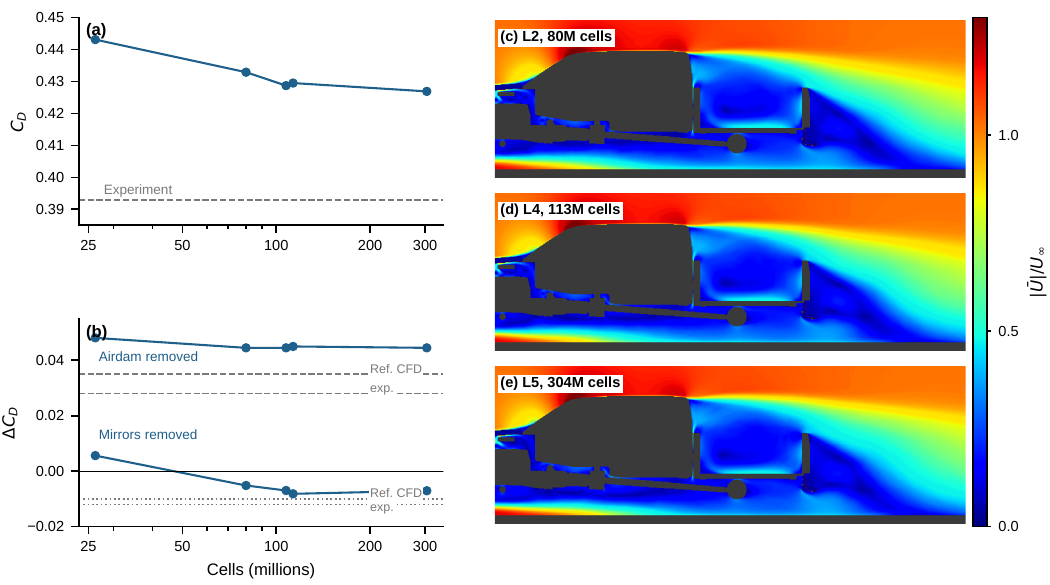}
\caption{Grid refinement on the baseline GTU. (a) Drag coefficient and (b) configuration deltas against cell count for the five-mesh family. Dashed and dotted lines are the experiment and the CFD of \citet{woodiga2020gtu}. (c)--(e) Time-averaged velocity magnitude on the centreplane for the 80-, 113- and 304-million-cell meshes.}
\label{fig:grid}
\end{figure}

The numerical setup was fixed in two steps on the baseline GTU, both run on the three configurations of Table~\ref{tab:validation}. A grid refinement study set the cell budget and the surface resolution of the body panels (Fig.~\ref{fig:grid}, Table~\ref{tab:grid}). The near-field meshing and the numerics were then tuned at that budget, with the two measured part-removal deltas as the acceptance test. The grid family has \nGridLevels{} meshes of \nGridCells{} million cells. They are built on a structured background mesh with volume refinement boxes and three prism layers, and they use the default limiter and blending of the solver. Surface refinement rises by one level from the first mesh to the second and stays fixed after that, so the third to fifth meshes differ only in volume refinement. Baseline drag falls from \nGridCDa{} on the coarsest mesh to \nGridCDe{} on the finest, and it changes by less than 0.003 from the third mesh onward. The configuration deltas converge on coarser meshes than the absolute level. The airdam delta lies between $+0.0445$ and $+0.0450$ on every mesh from 80 million cells. The mirror delta has the wrong sign on the coarsest mesh and settles between $-0.0070$ and $-0.0082$ from 107 million cells. The mean velocity on the centreplane (Fig.~\ref{fig:grid}c--e) is indistinguishable between meshes from 80 million cells upward.

\begin{table}[h]
\caption{Grid refinement study on the baseline GTU. Deltas are referenced to each mesh's own baseline.}
\label{tab:grid}
\centering
\small
\begin{tabular}{lrrrr}
\toprule
Mesh & Cells (M) & $C_D$ & $\Delta C_D$ mirrors & $\Delta C_D$ airdam \\
\midrule
L1 & 26 & 0.4431 & $+0.0056$ & $+0.0481$ \\
L2 & 80 & 0.4329 & $-0.0052$ & $+0.0445$ \\
L3 & 107 & 0.4287 & $-0.0070$ & $+0.0445$ \\
L4 & 113 & 0.4295 & $-0.0082$ & $+0.0450$ \\
L5 & 304 & 0.4269 & $-0.0071$ & $+0.0445$ \\
\midrule
Experiment \citep{woodiga2020gtu} & --- & 0.3930 & $-0.0120$ & $+0.0280$ \\
CFD \citep{woodiga2020gtu} & --- & --- & $-0.0100$ & $+0.0350$ \\
\bottomrule
\end{tabular}
\end{table}

\subsection{From the grid family to the dataset setup}
The cell budget was set at about 100 million cells, with the body panels at surface level 8. The near-field meshing follows the practice of the Automotive CFD Prediction Workshop \citep{hupertz2022autocfd2,economon2024autocfd4}. Body-conformal refinement shells and feature-edge refinement replace the volume boxes of the grid family. Dedicated boxes refine the mirror wakes, and the mirrors and the chin spoiler receive finer surface refinement (Appendix~\ref{app:cfd}). Five prism layers replace three, which removed a premature separation that the thinner stack had produced on the body panels. The slope limiter was removed and the low-dissipation blending of the convective flux was raised to suit the finer near field. A sweep of the blending over its usable range moves the mirror delta by about \nBlendSens{} and the airdam delta by less than 0.001. On the baseline GTU the dataset setup gives \nCellsBase{} million cells and the two deltas of Table~\ref{tab:validation}, both within 0.001 of the reference CFD. The absolute drag responds more strongly to these choices than the deltas. It moves by 0.02 between the grid family and the dataset setup, while the deltas move by 0.003 to 0.010. Neither study removes the offset in absolute drag from the experiment. The limiter and blending were set with the two experimental deltas in view, and the sweep above bounds how far that choice moves each delta.

\subsection{Repeatability}
Some cases of the dataset were simulated a second time with unchanged shape parameters. Across these repeated pairs, drag agrees to within \nRepeatCDLo{} to \nRepeatCDHi{}\%.

\subsection{Comparison with experiment and reference CFD}
\begin{figure}[h]
\centering
\includegraphics[width=\linewidth]{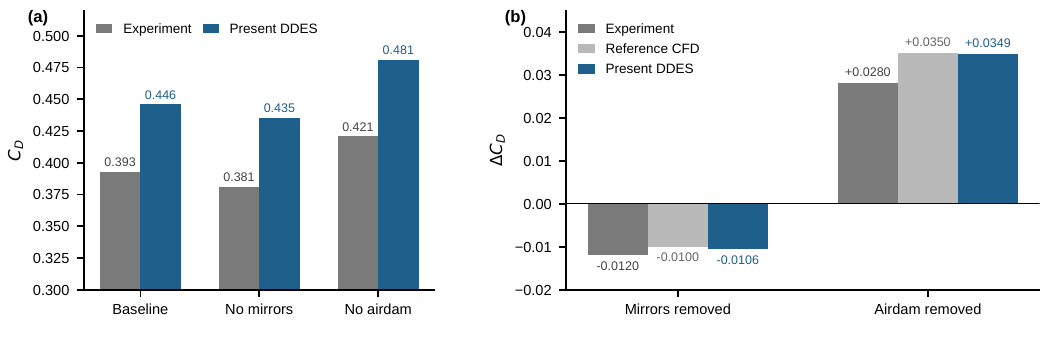}
\caption{Validation on the baseline GTU. (a) Absolute drag coefficient for the three configurations, from the experiment of \citet{woodiga2020gtu} and the present DDES. (b) Drag change on removing the mirrors and the airdam, from the experiment and CFD of \citet{woodiga2020gtu} and the present DDES.}
\label{fig:validation}
\end{figure}

We compare the GTU baseline and two single-part removals, the mirrors and the airdam, with the wind-tunnel measurements and the CFD of \citet{woodiga2020gtu} for the same configurations (Fig.~\ref{fig:validation}, Table~\ref{tab:validation}). The measurements were taken in a three-quarter open-jet wind tunnel with a five-belt rolling road at \nUinfKmh{}~km\,h$^{-1}$. The coefficients use the same frontal area of \nRefArea{}~m$^2$ as ours. The reference CFD is an SA-DDES on an unstructured mesh of about 100 million cells at $\yplus$ of about 30. It uses a time step of $2\times10^{-4}$~s, averages over the final 5.5~s of a 6~s run and uses a closed domain of the same size as ours. It is a simulation of comparable resolution and modelling with an independent solver. The source reports the reference CFD as deltas only, and we compare the absolute level with the experiment alone.

The configuration deltas agree with experiment on both removals. Removing the mirrors reduces drag by 0.0106 in the present simulations, against 0.012 measured and 0.010 in the reference CFD. Removing the airdam raises drag by 0.0349, against 0.028 measured and 0.035 in the reference CFD. Both signs are correct. The mirror delta lies between the two reference values. The airdam delta matches the reference CFD to within 0.0001 and exceeds the measurement by \nValTol{}. The two removals are independent and of opposite sign. The mirrors are small appendages in locally attached flow, and the airdam controls the flow into the underbody. Agreement on two such different effects makes a single cancellation of errors unlikely. We keep the label ``no airdam'' of \citet{woodiga2020gtu} for the second removal. The part is the baseline airdam of the GTU and is distinct from the chin spoiler of Appendix~\ref{app:geometry}.

The absolute level agrees less well. The present baseline drag coefficient is \nBaseCD{} against a measured \nExpBaseCD{}, an offset of \nAbsOffsetCD{}. An offset of the same sign appears on every mesh and every setting of the grid study, including the finest mesh, so resolution does not explain it. The experiment used an open-jet tunnel with a nozzle area of 22.45~m$^2$, a nozzle blockage of 12\%. \citet{woodiga2020gtu} state no blockage correction, Reynolds number or reference length. Both simulations, ours and the reference CFD, use a closed domain with a blockage of \nBlockage{}\%. The absolute level of the reference CFD is not reported, and we cannot tell whether it carries a similar offset. The open jet against the closed domain, the reference convention and the wall-modelled near-wall treatment are the candidate contributions, and the present data cannot separate them. \citet{woodiga2020gtu} use a domain of the same kind, note that it omits the nozzle, supports and collector of the tunnel, and expect differences from the experiment for that reason. Their simulation over-predicts the airdam delta by the same amount as ours. Comparisons of absolute drag from this dataset with tunnel values should carry this offset.

%% file: appendix/D-design-space.tex
\section{Aerodynamics of the design space}
\label{app:aero}

\begin{figure}[h]
\centering
\includegraphics[width=\linewidth]{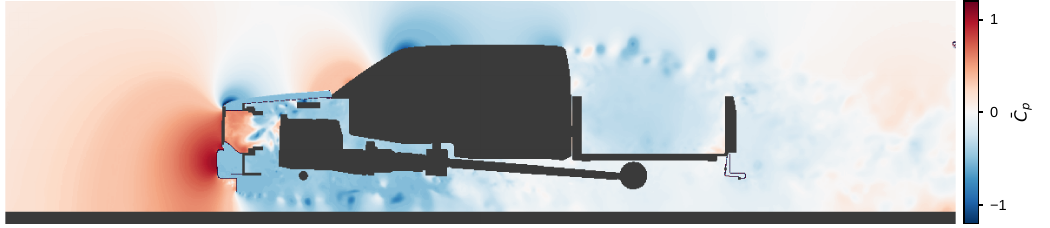}
\caption{Time-averaged pressure coefficient on the centreplane $y=0$ of the baseline GTU, the same plane and case as Fig.~\ref{fig:baseflow}.}
\label{fig:basecp}
\end{figure}

\subsection{Baseline flow}
The flow accelerates over the hood leading edge and again over the roof leading edge. A thin shear layer leaves the A-pillar and the roof and stays attached along the cab. Behind the cab the flow separates from the roof trailing edge and the cab back. A low-momentum recirculation fills the cargo bed between the cab back, the bed floor and the tailgate. The tailgate sheds a second separation into the base wake, and the bed flow feeds this wake from above. Below the vehicle the flow accelerates between the floor and the underbody into a jet that leaves behind the rear axle and bounds the base wake from below. The mean pressure (Fig.~\ref{fig:basecp}) shows stagnation on the front face and the airdam, suction over the hood and roof leading edges, and a nearly uniform base pressure over the tailgate and the bed. In the instantaneous field (Fig.~\ref{fig:baseflow}b) every separated region is resolved as unsteady turbulence, and the modelled boundary layer is confined to the attached surfaces.

\subsection{Drag distribution}
Over the \nCorpus{} cases of the benchmark set, drag spans 0.277 to 0.549 with a standard deviation of \nCDSd{} (Fig.~\ref{fig:designspace}a). The median drag is \nCDMedClosed{} for closed-bed cases and \nCDMedOpen{} for open-bed cases, with standard deviations of 0.045 and 0.041. Between the two bed states the median shifts by about one standard deviation, and the spread stays about the same. Drag rank-correlates most strongly with overall height, with a Spearman coefficient of $\nRhoHeightClosed$ for closed beds and $\nRhoHeightOpen$ for open beds. Its rank correlation with body length is weak, at $-0.14$ and $-0.05$.

\subsection{Bed state}
\nBedPairsMatched{} base trucks carry both bed states with the chin spoiler present in both. Opening the bed raises drag on every one of them. The median increase is $\nBedMedian$, and the increases range from $\nBedMin$ to $\nBedMax$ (Fig.~\ref{fig:designspace}b).

\subsection{Chin spoiler}
\nSpoilerPairs{} base trucks carry both spoiler states with the bed closed. Every open-bed case carries the spoiler. Adding the spoiler raises drag on every one of them, with a median increase of \nSpoilerMedian{}.

%% file: appendix/E-benchmark-details.tex
\section{Benchmark models and predictions}
\label{app:benchmark}

\subsection{Architectures}

Each model maps a query point on the surface or in the volume, together with a representation of the geometry, to the field values at that point. All four were trained from scratch with the architecture and optimizer of their reference implementations. The training settings and model-specific losses are given below.

\paragraph{\DoMINO{}.}
\DoMINO{} \citep{domino2025} is a point-cloud model. It predicts the fields at each output point from the geometry in a local region around that point, combined with a coarse encoding of the whole body. Working on local regions lets it run at the full resolution of the mesh.

\paragraph{\GeoT{}.}
\GeoT{} \citep{geotransolver2025} extends the Transolver backbone. Self-attention acts on learned slices of the flow state, and every block cross-attends to a shared context built from multi-scale ball queries on the geometry and from the operating parameters. The geometry enters every layer of the network through this context.

\paragraph{\ABUPT{}.}
\ABUPT{} \citep{abupt2025} is an anchored-branched universal physics transformer. A fixed set of anchor tokens on the surface and in the volume carries the attention, and each query point is decoded from the anchors by cross-attention. The cost of attention is independent of the number of query points at inference.

\paragraph{\SMART{}.}
\SMART{} \citep{smart2026} is a transformer that encodes a point cloud of the geometry and the simulation parameters into a shared latent space. A physics decoder maps query points to field values by cross-attending to the intermediate latents of the encoder. The geometric features and the predicted field are updated together through the decoder.

\subsection{Training configuration}
\label{app:training}

Every model is trained for \nTrainEpochs{} epochs in FP32 in both tracks. Table~\ref{tab:training} gives the optimizer, learning-rate schedule, geometry and query sampling budgets, target normalization and loss for each model. The surface-track runs use eight NVIDIA H100 GPUs each. \GeoT{} uses Muon for two-dimensional weight tensors and AdamW for the remaining parameters. Lion uses $\beta=(0.9,0.99)$ for \ABUPT{} and \SMART{}, and Adam uses $\beta=(0.9,0.999)$ for \DoMINO{}. \ABUPT{} excludes normalization parameters and biases from weight decay. Its warmup lasts 5\% of training, or 25 epochs. \DoMINO{} redraws its surface training queries each epoch. \ABUPT{} pools its geometry points to 16,384 supernodes.

\GeoT{} and \DoMINO{} use mean squared error (MSE), \ABUPT{} uses $L_1$ loss, and \SMART{} uses relative $L_2$ loss. For volume-track training, \GeoT{} and \SMART{} average the surface and volume loss terms. \SMART{} additionally weights near-wall and wake regions in its volume loss during training. The common evaluation metrics are those of \S\ref{sec:benchmark}, and every model is evaluated on the full volume mesh. The \SMART{} volume model has 12.8 million trainable parameters, as reported in Table~\ref{tab:results}.

% Final checkpoint-selection rule, batch sizes, normalization-statistics
% procedure, exact SMART spatial loss weights, and implementation versions
% have not yet been supplied; do not infer them from the superseded run ledger.
\input{tables/t_training}

\subsection{Surface predictions}
\label{app:surface}

\begin{figure}[p]
\centering
\includegraphics[width=\linewidth]{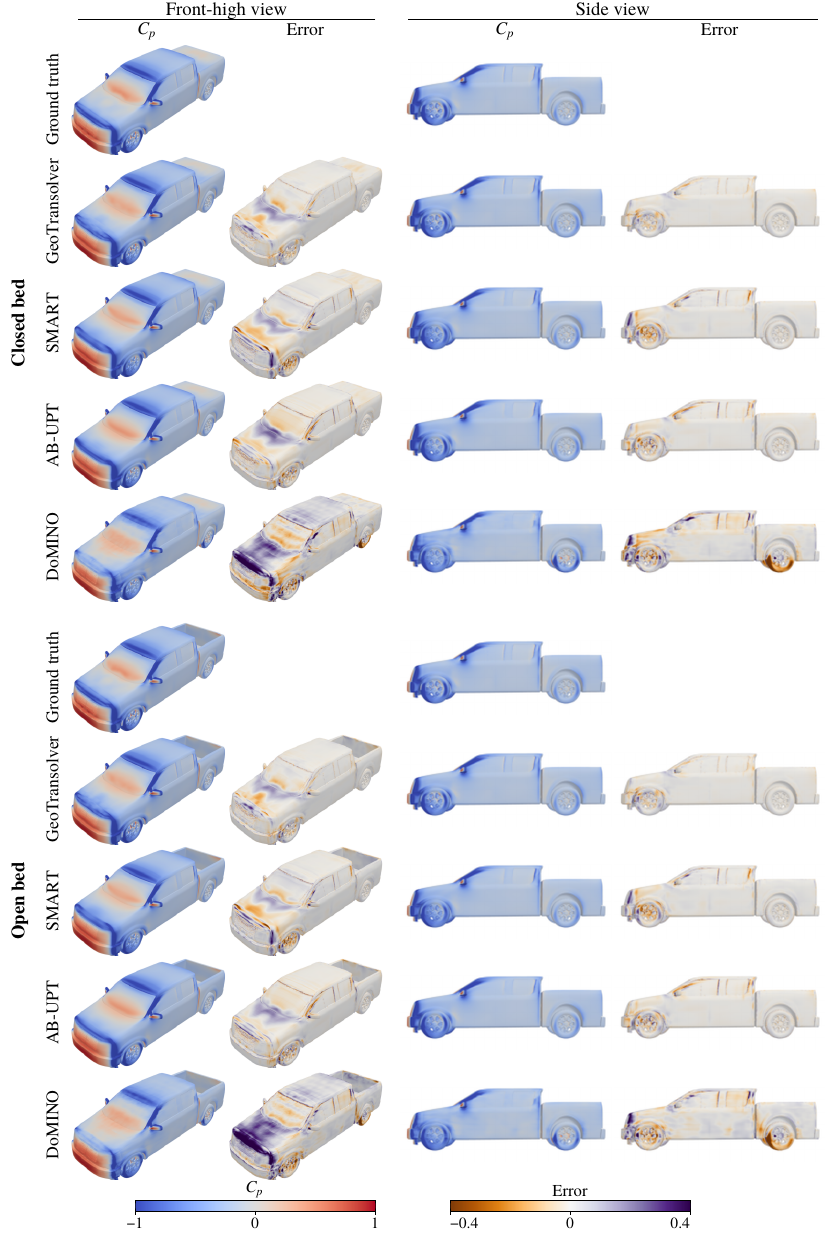}
\caption{Surface pressure coefficient for the matched pair of Fig.~\ref{fig:fields}, with the bed closed (top) and open (bottom), in the front-high and side views. Each prediction is shown beside its error, the prediction minus the ground truth. Both bed states share one error scale.}
\label{fig:surfcp}
\end{figure}

\begin{figure}[p]
\centering
\includegraphics[width=\linewidth]{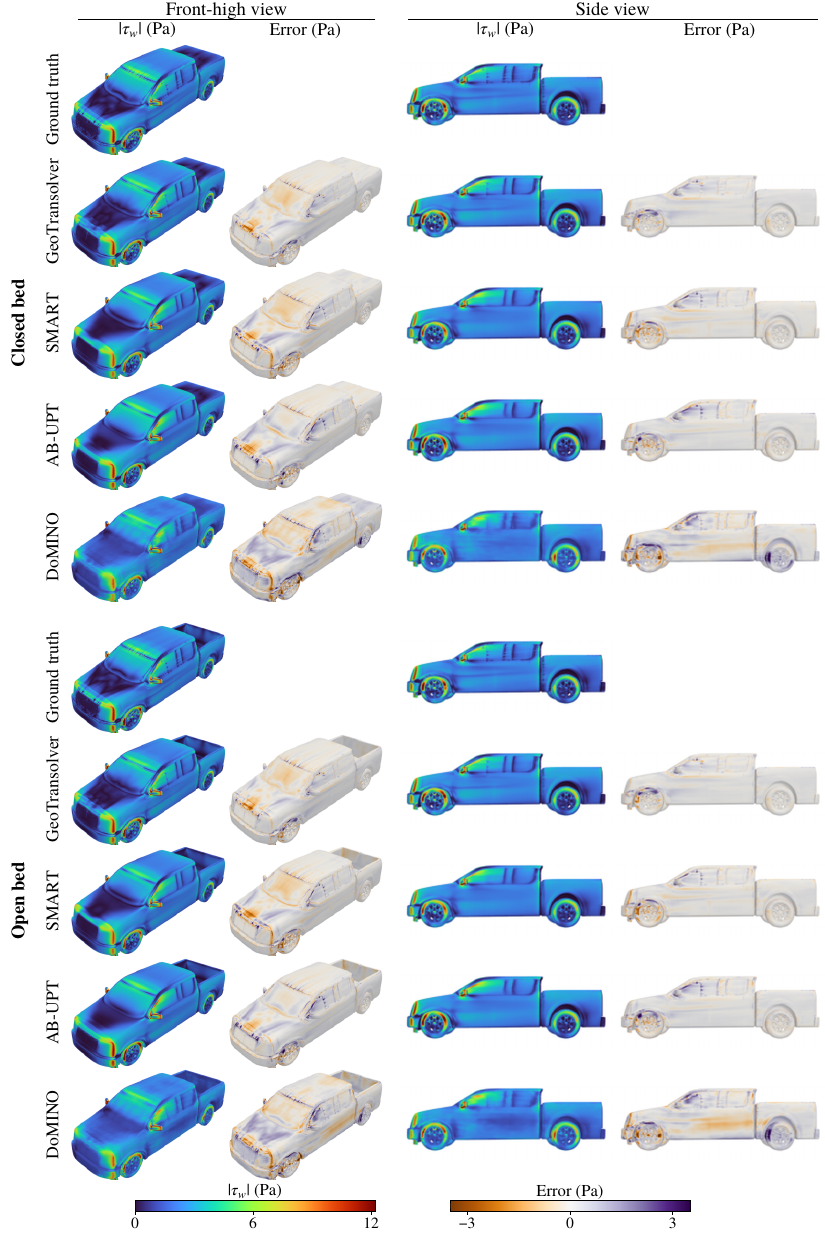}
\caption{Wall-shear-stress magnitude for the matched pair of Fig.~\ref{fig:fields}, laid out as in Fig.~\ref{fig:surfcp}.}
\label{fig:surfwss}
\end{figure}

Figures~\ref{fig:surfcp} and~\ref{fig:surfwss} give the surface pressure coefficient and the wall-shear-stress magnitude for the matched pair of Fig.~\ref{fig:fields}, in the front-high view of the main text and in a side view. Each prediction is shown beside its error, the prediction minus the ground truth. The error colour scale is symmetric at the 99th percentile of the absolute error over the four models. The two bed states give the same value to within 1.5\%, and both share one scale.

\paragraph{Pressure.}
The pressure error of every model sits on the front of the truck. It is largest on the grille and the front fascia, along the hood leading edge, and in a band across the hood where the pressure recovers toward the windshield base. The wheels and wheel arches carry a second concentration of error. The cab sides, the roof and the bed sides carry little error. \GeoT{} has the weakest band across the hood. \SMART{} and \ABUPT{} carry a band of alternating sign across the hood, where the pressure rises toward the windshield. \DoMINO{} over-predicts the pressure over the grille, the fascia and the hood leading edge, and under-predicts it on the rear wheel. Opening the bed leaves the error on the front half of the truck unchanged.

\paragraph{Shear stress.}
The shear-stress error concentrates on the hood, the grille and the wheels. \GeoT{}, \SMART{} and \ABUPT{} under-predict the shear over the centre of the hood, where the true field carries streaks of low shear. The under-prediction is weakest for \GeoT{} and strongest for \SMART{}. \DoMINO{} over-predicts the shear where the true shear is low, on the hood, the cab sides and the bed floor. Its side view also shows a band of under-predicted shear along the lower body side.

\subsection{Volume predictions}
\label{app:volume}

\begin{figure}[p]
\centering
\includegraphics[width=\linewidth]{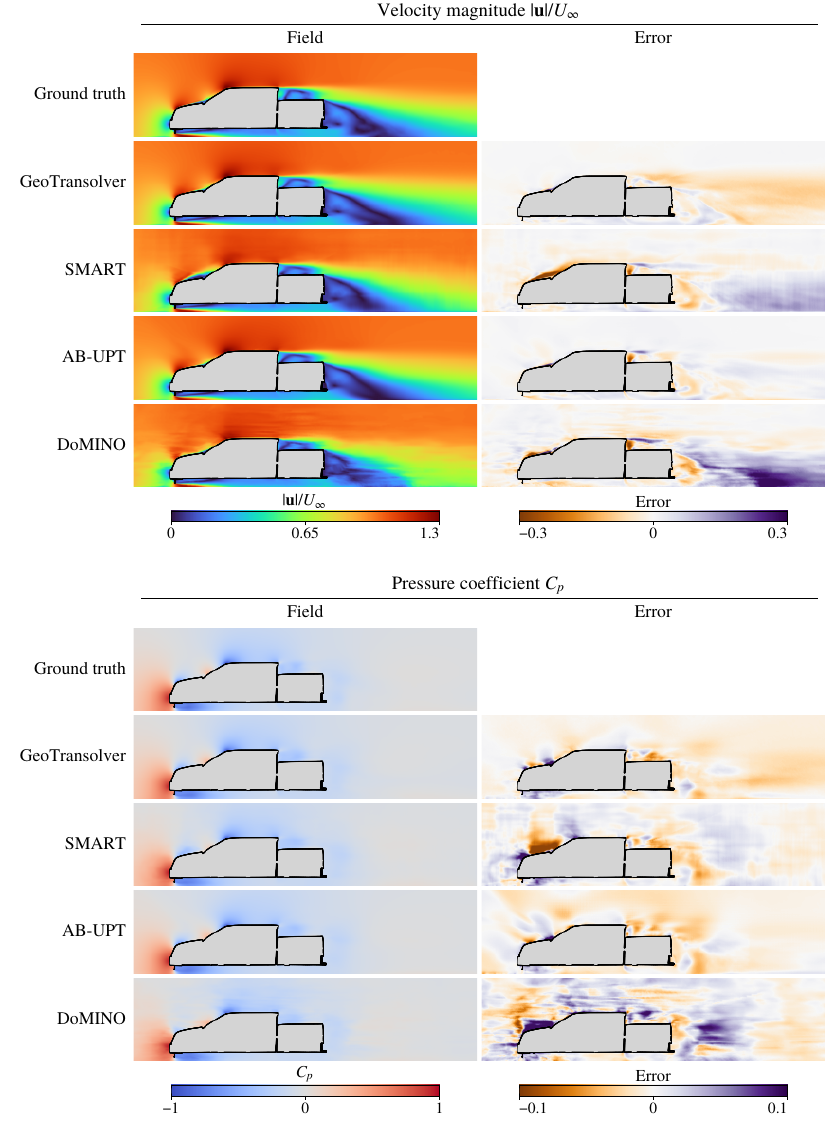}
\caption{Velocity magnitude (top) and pressure coefficient (bottom) on the centreplane $y=0$ for the closed-bed truck of the matched pair of Fig.~\ref{fig:fields}. Each prediction is shown beside its error, the prediction minus the ground truth. Figure~\ref{fig:volopen} uses the same colour scales.}
\label{fig:volclosed}
\end{figure}

\begin{figure}[p]
\centering
\includegraphics[width=\linewidth]{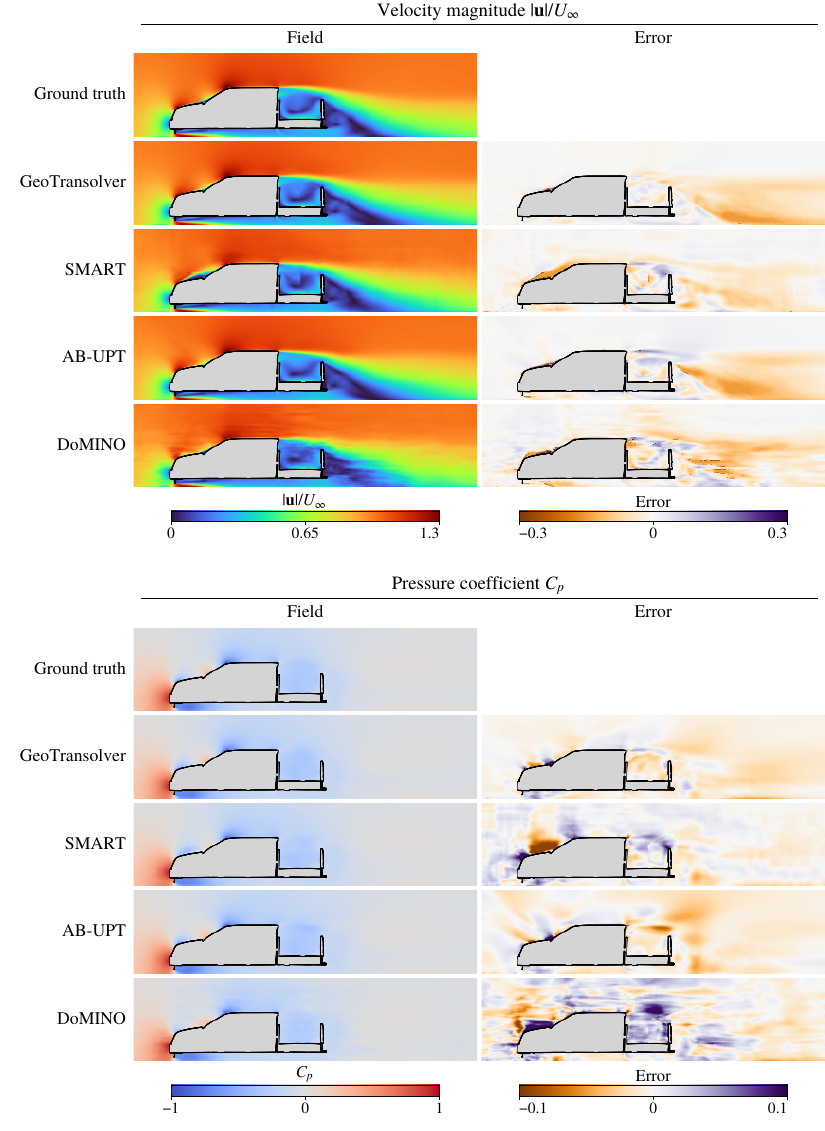}
\caption{Velocity magnitude (top) and pressure coefficient (bottom) on the centreplane $y=0$ for the open-bed truck of the matched pair, laid out as in Fig.~\ref{fig:volclosed}.}
\label{fig:volopen}
\end{figure}

Figures~\ref{fig:volclosed} and~\ref{fig:volopen} give the velocity magnitude and the pressure coefficient on the centreplane $y=0$ for the same pair. The ground truth is the released field at the native cell centres that lie within a thin slab around the plane. Each model is queried at the same cell centres, mixed with a random background drawn from the rest of the volume, and only the slab points are read out. The points are triangulated in the plane and interpolated to a 5\,mm lattice for display. \DoMINO{} returns no value outside its bounding box, and those points are left blank. The error colour scale is symmetric and common to both bed states, and it covers the 99th percentile of the absolute error over the four models.

\paragraph{Velocity.}
The velocity error sits in the shear layer that leaves the roof trailing edge, in the bed and in the near wake. With the bed closed, \GeoT{} and \ABUPT{} carry small errors at the roof trailing edge and in the upper wake. \SMART{} under-predicts the velocity in a thin layer over the hood and the windshield. \DoMINO{} over-predicts the velocity in the lower wake, and its error is noisy through the outer flow. With the bed open, every model carries error inside the bed recirculation and along the shear layer above it. \GeoT{} and \ABUPT{} under-predict the velocity through the near and far wake.

\paragraph{Pressure.}
The pressure error is largest around the front bumper and at the base of the windshield, where \SMART{} under-predicts the pressure. \GeoT{}, \SMART{} and \ABUPT{} predict a smooth field with a small negative offset through the wake. \DoMINO{} carries a noisy error through the outer flow, where the true field is smooth. \DoMINO{} also over-predicts the pressure in the near wake of the closed truck.

%% file: tables/t_training.tex
% Source: author-provided training ledger, with corrections confirmed
% 2026-09-26: 500 epochs and FP32 for every run, full-mesh volume evaluation
% for GeoTransolver as well as the other models, and 12.8M SMART volume parameters.
% Superseded checkpoint indices, clock-based seeds and volume GPU counts
% are intentionally not carried over from the ledger.
\begin{table}[htbp]
\caption{Training configurations. All models use \nTrainEpochs{} epochs in FP32. Query counts are per case during training; volume evaluation uses the full mesh for every model. The volume-track rows give the surface and volume query counts used together in that track.}
\label{tab:training}
\centering
\small
\setlength{\tabcolsep}{3pt}
\renewcommand{\arraystretch}{1.2}
\begin{tabular}{@{}p{0.21\linewidth}p{0.19\linewidth}p{0.17\linewidth}p{0.17\linewidth}p{0.18\linewidth}@{}}
\toprule
Setting & \GeoT{} & \ABUPT{} & \DoMINO{} & \SMART{} \\
\midrule
Optimizer & Muon / AdamW & Lion & Adam & Lion \\
Base learning rate & $10^{-3}$ & $5\times10^{-5}$ & $10^{-3}$ & $5\times10^{-5}$ \\
Weight decay & $10^{-4}$ & $0.05$ & $0$ & $0$ \\
Learning-rate schedule & Halve every 100 epochs; no warmup & 5\% linear warmup; cosine to $10^{-6}$ & Halve at epochs 50, 100, 200, 250, 300, 350, 400, 450 & Cosine to $0$ \\
Gradient clipping & None & Norm $2.0$ & Norm $10$ & Norm $2.0$ \\
Geometry points & 200,000 & 200,000 & 300,000 & 200,000 \\
Target normalization & Mean/std & Mean/std & Min--max & Mean/std \\
Field loss & MSE & $L_1$ & MSE & Relative $L_2$ \\
\midrule
Surface-track queries & 60,000 & 16,384 & 32,000 & 16,000 \\
Volume-track surface queries & 16,000 & 16,384 & 16,000 & 16,000 \\
Volume-track volume queries & 80,000 & 16,384 & 32,000 & 80,000 \\
\bottomrule
\end{tabular}
\end{table}

%% file: appendix/F-release.tex
\section{Release, files and datasheet}
\label{app:release}

\subsection{Access and license}
\ifdefined\ArxivVersion The dataset is released under the CC-BY-NC-4.0 license and is available at \ShiftTruckURL. Access is gated: it requires a Hugging Face account and an access request, which shares the user's contact information with the maintainers.\else The dataset is released under the CC-BY-NC-4.0 license, and is available with its Croissant metadata \citep{akhtar2024croissant} at \texttt{gs://shift-truck-sample/}.\fi{} The GTU baseline remains under the terms of its distributor, ECARA, and work that uses it should cite \citet{woodiga2020gtu}.

\subsection{Files}
Each case is one directory of six files (Table~\ref{tab:files}). The root of the release holds \texttt{doe\_parameters.csv}, with the design parameters, force coefficients and provenance of every case. It also holds the license and three utility scripts, which check the release, integrate the surface fields to forces and visualize a case. A case takes about 781~MB without its volume file and 11.9~GB with it. \ifdefined\ArxivVersion The full release takes about 11.9~TB. The surface, force and metadata files of all cases take about 0.8~TB\else The full release takes about 9.5~TB. The surface, force and metadata files of all cases take about 0.6~TB\fi{} and can be downloaded without the volume files, which suffices for the surface track.

\begin{table}[h]
\caption{Files per case, with sizes for \texttt{variant\_0000}.}
\label{tab:files}
\centering
\small
\resizebox{\linewidth}{!}{%
\begin{tabular}{llll}
\toprule
File & Format & Contents & Size \\
\midrule
\texttt{merged\_surfaces.stl} & binary STL & vehicle surface, triangulated & 386 MB \\
\texttt{merged\_surfaces.vtp} & VTK PolyData & time-averaged pressure and wall shear stress, point and cell data & 395 MB \\
\texttt{merged\_volumes.vtu} & VTK UnstructuredGrid & time-averaged pressure and velocity on the full mesh & 11.1 GB \\
\texttt{forces.json} & JSON & $C_D$, $C_L$, length of the averaging window & --- \\
\texttt{metadata.json} & JSON & solver run provenance, iteration count, physical time, removed patches & --- \\
\texttt{params.json} & JSON & reference values of Table~\ref{tab:refquant} & --- \\
\bottomrule
\end{tabular}}
\end{table}

\subsection{Fields}
The surface file carries \texttt{Pressure Average (Pa)}, the time-averaged absolute static pressure, and the three components of \texttt{Wall Shear Stress Average (N/m\textsuperscript{2})}. Both are stored as point data and as cell data. The file also carries the unit normals and the index of each point and cell in the full simulation surface. The volume file carries \texttt{Pressure Average (Pa)} and the three components of \texttt{Velocity Average (m/s)}. Gauge pressure follows from subtracting the reference pressure of Table~\ref{tab:refquant}. The contact patches under the wheels are open, and the absolute pressure level does not cancel in surface integrals. The surface covers the vehicle alone. The tunnel walls, the inlet, the outlet and the floor patches are removed, and \texttt{metadata.json} lists them. Every field and force coefficient is a time average over the final \nAvgStepsCorpus{} steps, and the release contains no instantaneous fields or force histories. Each case has its own mesh, with 3.6 to 4.0 million surface points and about 98 million volume cells.

\subsection{Benchmark split}
\label{app:split}
\ifdefined\ArxivVersion The case lists of the split of \S\ref{sec:benchmark} are available from the authors on request.\else The three case lists of the split of \S\ref{sec:benchmark} are released with the dataset.\fi{} In the 15-dimensional parameter space, with each parameter normalized to $[-1,1]$, the median distance from a test truck to its nearest training truck is \nSplitNNTest{}, against \nSplitNNTrain{} from each training truck to its nearest other training truck, and the two distributions do not differ (two-sample Kolmogorov--Smirnov test, $p=\nSplitNNp{}$). None of the \nParamsCont{} parameter distributions differs between the training and test sets ($p \geq \nSplitMargp{}$). Test drag spans the range of the benchmark set, from \nSplitCDTestLo{} to \nSplitCDTestHi{}, with a median of \nSplitCDTestMed{} against \nSplitCDTrainMed{} in training.

\subsection{Maintenance}
\ifdefined\ArxivVersion Corrections and additions to the release are recorded in the change log of the dataset card.\else Cases are added to the release as their simulations complete. Each addition is recorded in a change log distributed with the release.\fi

\subsection{Datasheet}
Table~\ref{tab:datasheet} summarizes the dataset under the headings of \citet{gebru2021datasheets}.

\begin{table}[t]
\caption{Datasheet.}
\label{tab:datasheet}
\centering
\small
\begin{tabular}{p{0.2\linewidth}p{0.74\linewidth}}
\toprule
Motivation & Training and benchmarking of neural surrogates for the external aerodynamics of pickup trucks, a vehicle class absent from open datasets. \\
Composition & \nRelease{} cases, each a morphed GTU pickup at one operating point, with \nParamsCont{} continuous shape parameters and two parts switched on and off. Each case has a surface with time-averaged pressure and wall shear stress, a volume with time-averaged pressure and velocity, and its drag and lift coefficients. \ifdefined\ArxivVersion The out-of-distribution cases of \S\ref{sec:ood} are not included. \fi The data contain no personal information. \\
Collection & Geometries morphed from GTU configuration 7 with a deformation cage (Appendix~\ref{app:geometry}), then meshed and simulated with DDES (Appendix~\ref{app:cfd}) on \nGPUsPerCase{} H100 GPUs per case. Appendix~\ref{app:vv} validates the setup. \\
Preprocessing & Fields and forces averaged over the final \nAvgStepsCorpus{} steps. Surface restricted to the vehicle. Instantaneous data discarded. Forces of two cases recovered by integration. \\
Uses & Surrogate training and benchmarking, studies of shape and drag, and transfer between vehicle classes. Absolute drag carries an offset against the wind tunnel (Appendix~\ref{app:vv}). The data do not replace simulation or testing in safety-relevant decisions. \\
Distribution & \ifdefined\ArxivVersion \raggedright Hugging Face, \ShiftTruckURL, with gated access on request. CC-BY-NC-4.0.\else Anonymized sample for review in a public cloud bucket, full release linked in the final version. CC-BY-NC-4.0, with Croissant metadata.\fi \tabularnewline
Maintenance & \ifdefined\ArxivVersion Corrections and additions recorded in the change log of the dataset card.\else Cases added as they complete and recorded in a change log distributed with the release.\fi \\
\bottomrule
\end{tabular}
\end{table}

%% file: appendix/G-datasets.tex
\section{Existing datasets}
\label{app:datasets}

Table~\ref{tab:landscape} lists the open high-fidelity datasets for external vehicle aerodynamics. AhmedML, WindsorML and DrivAerML \citep{ashton2024ahmedml,ashton2024windsorml,ashton2024drivaerml} publish scale-resolving solutions, from detached-eddy or wall-modelled large-eddy simulation, on reference bodies, with a few hundred cases each. DrivAerNet, DrivAerNet++ \citep{elrefaie2024drivaernet,elrefaie2024drivaernetpp} and DrivAerStar \citep{qiu2025drivaerstar} extend a DrivAer-derived design space to thousands of cases with steady RANS. SHIFT-SUV \citep{shift_placeholder} applies detached-eddy simulation to parametric variants of the AeroSUV. \SHIFTTruck{} is the only entry that describes a pickup truck. Every dataset validates on one to four baseline geometries and none on its morphed variants.

\begin{table}[h]
\caption{Open high-fidelity datasets for external vehicle aerodynamics. \textit{n/s}: not stated in the source. Case counts are those of each release as of September 2026. Validation gives the drag error against the wind tunnel on the baseline geometries; where a source gives coefficients only, the error is computed from them.}
\label{tab:landscape}
\centering
\small
\resizebox{\linewidth}{!}{%
\begin{tabular}{llrlccl}
\toprule
Dataset & Class & Cases & Method & Surface & Volume & Experimental validation \\
\midrule
AhmedML & bluff body & \nLandAhmedML{} & SA-DDES & yes & yes & $C_D$ 4.7\%, $C_L$ 0.6\% (baseline) \\
WindsorML & bluff body & \nLandWindsorML{} & WMLES & yes & yes & $C_D$ 0.34 vs.\ 0.33 (baseline); PIV \\
DrivAerML & passenger car & \nLandDrivAerML{} & SA-DDES & yes & yes & $C_D$ 7.5\% (baseline) \\
DrivAerNet & passenger car & \nLandDrivAerNet{} & RANS, $k$--$\omega$ SST & yes & yes & $C_D$ 0.81\% (baseline) \\
DrivAerNet++ & passenger car & \nLandDrivAerNetPP{} & RANS, $k$--$\omega$ SST & yes & yes & $C_D$ 2.2--4.9\% (4 baselines) \\
DrivAerStar & passenger car & \nLandDrivAerStar{} & RANS, $k$--$\omega$ SST & yes & yes & $C_D$ 1.04\% (mean, 3 baselines) \\
SHIFT-SUV & SUV & \nLandSHIFTSUV{} & DDES & yes & yes & n/s (FKFS tunnel) \\
\textbf{\SHIFTTruck{}} & \textbf{pickup truck} & \nRelease{} & SA-DDES & yes & yes & $C_D$ 13.5\% (baseline); $\Delta C_D$ within 0.007 \\
\bottomrule
\end{tabular}}
\end{table}